\documentclass[prd,aps,twocolumn]{revtex4}
\usepackage{graphicx,url}

\usepackage[sort&compress]{natbib}
\usepackage[utf8]{inputenc}
\def\gr{$\gamma$-ray}
\begin{document}

\title{Multi-messenger gamma-ray counterpart of the IceCube neutrino signal} 

\author{A.~Neronov$^{1}$}
\author{M.~Kachelrie\ss$^{2}$}
\author{D.~V.~Semikoz$^{3,4}$}

\affiliation{$^1$Astronomy Department, University of Geneva,
Ch. d'Ecogia 16, Versoix, 1290, Switzerland}
\affiliation{$^2$Institutt for fysikk, NTNU, Trondheim, Norway}
\affiliation{$^{3}$APC, Universite Paris Diderot, CNRS/IN2P3, CEA/IRFU,
Observatoire de Paris, Sorbonne Paris Cite, 119 75205 Paris, France}
\affiliation{$^{4}$National Research Nuclear University MEPHI (Moscow Engineering Physics Institute), Kashirskoe highway 31, 115409 Moscow, Russia}

\begin{abstract}
A signal of high-energy extraterrestrial neutrinos from unknown source(s)
was recently discovered by the IceCube experiment. Neutrinos are always
produced together with $\gamma$-rays, but the $\gamma$-ray flux from
extragalactic sources is suppressed due to attenuation in the intergalactic
medium. We report the discovery of a $\gamma$-ray excess at high Galactic
latitudes starting at energies 300\,GeV in the data of the Fermi telescope.
We show that the multi-TeV $\gamma$-ray diffuse emission has spectral
characteristics at both low and  high Galactic latitudes compatible with
those of the  IceCube high neutrino signal in the same sky regions.
This suggests that these $\gamma$-rays are the counterpart of the IceCube
neutrino signal, implying that a sizable part of the IceCube neutrino flux
originates from the Milky Way. 
We argue that the diffuse neutrino and $\gamma$-ray signal at high Galactic
latitudes originates either from  previously unknown nearby cosmic ray
"PeVatron" source(s), an extended Galactic CR halo or from decays of heavy 
dark matter particles. 
\end{abstract}

\maketitle

\section{Introduction}%
The discovery of an extraterrestrial neutrino signal in the TeV--PeV energy range by the IceCube collaboration \cite{icecube_science,icecube_pev} has recently opened the era of multi-messenger astronomy. High-energy neutrinos are produced by cosmic rays (CR) interacting at their acceleration sites or during propagation through interstellar and intergalactic space \cite{review1}. Alternatively, neutrinos may be produced in decays of metastable heavy dark matter (DM) particles~\cite{Berezinsky:1991aa,Gondolo:1991rn}. 
The source(s) of this  neutrino signal have remained unidentified so far because of the limited statistics of the IceCube data. Moreover, the High-Energy Starting Events (HESE) which provide the most significant contribution to the neutrino signal have a poor angular resolution. At the same time, the production of high-energy neutrinos is accompanied by $\gamma$-rays.  This implies that the neutrino sources could be identified using a "multi-messenger" approach by combining neutrino and $\gamma$-ray data \cite{review2}. 

The TeV--PeV $\gamma$-ray flux from distant sources is suppressed by  electron-positron pair production in interactions with low-energy photons of the extragalactic background light and the cosmic microwave background \cite{franceschini}.   Therefore, the presence or absence of a $\gamma$-ray counterpart can be used to clarify the origin of the neutrino signal: If the signal originates from extragalactic sources at cosmological distances, no $\gamma$-ray counterpart is expected in the multi-TeV to PeV band. In contrast, a Galactic origin  implies the presence of a comparable multi-TeV $\gamma$-ray flux. 

The search for the $\gamma$-ray counterpart of the  neutrino signal is challenging with both ground and space-based $\gamma$-ray telescopes. Ground-based telescopes like HESS~\footnote{\protect\url{https://www.mpi-hd.mpg.de/hfm/HESS/}}, MAGIC~\footnote{\protect\url{https://magic.mpp.mpg.de}} and VERITAS~\footnote{\protect\url{https://veritas.sao.arizona.edu}} or air shower arrays like HAWC~\footnote{\protect\url{https://www.hawc-observatory.org}} and ARGO-YBJ~\footnote{\protect\url{http://argo.na.infn.it}} suffer from a high background of events produced by charged CRs \cite{gound-based}. The arrival directions of the CR background events are distributed over large angular scales, similar to the expected $\gamma$-ray counterpart of the neutrino signal. Space-based telescopes like the Fermi Large Area Telescope (LAT) \cite{atwood09} achieve a much better suppression of the charged CR background, but they have small collection areas which severely limit the signal statistics.   

In this {\em Letter\/}, we report a study of the TeV diffuse gamma-ray sky based on the data of Fermi/LAT. The small effective area of Fermi/LAT is compensated by the very long exposure time of nine years of the Fermi/LAT data we use. We show that the  \gr\ flux and spectrum at low and high  Galactic latitudes are compatible with the flux of the measured neutrino signal, in the energy range where the two signals overlap. We suggest that the $\gamma$-ray in the multi-TeV band  is the counterpart of the IceCube neutrino signal.

\section{Cross-calibration of the LAT data in the multi-TeV band}%
Our analysis uses  events from the ULRACLEANVETO class collected by Fermi/LAT  during the period
between October 28, 2008 and December 15, 2017. We calculate the spectra of large sky regions using the  "aperture photometry" approach~\footnote{\protect\url{https://fermi.gsfc.nasa.gov/ssc/data/analysis/scitools/aperture_photometry.html}}. 

The energy resolution and the calibration of the telescope effective area degrade in the TeV band  \cite{bruel12,pass8}. Therefore we perform an additional cross-calibration of the Fermi/LAT flux measurements with those of ground-based $\gamma$-ray telescopes via a comparison of spectral measurements of the stacked spectra of selected calibration sources, see the Appendix for details.  We find that  a cross-calibration factor $\kappa=1-c\log\left(E/100\mbox{ GeV}\right)$, with $c=0.25\pm 0.12$  has to be applied to the LAT flux measurements above  $300$~GeV to achieve better consistency with the ground-based telescope measurements. We apply this factor in our analysis, following a practice common in X-ray data analysis~\footnote{\protect\url{https://heasarc.gsfc.nasa.gov/docs/heasarc/caldb/caldb_xcal.html}}. The uncertainty of the parameter $c$  is taken into account as an additional systematic error. We have verified that the cross-calibration factor also assures the consistency of the Fermi/LAT measurements of diffuse TeV emission from large regions of the sky with the measurements by the ground-based air shower arrays ARGO-YBJ \cite{argo} and MILAGRO \cite{milagro}, see the Appendix for details.

\section{Diffuse TeV \gr\ signal}%
\begin{figure}
  \vskip-0.5cm
\includegraphics[width=\linewidth]{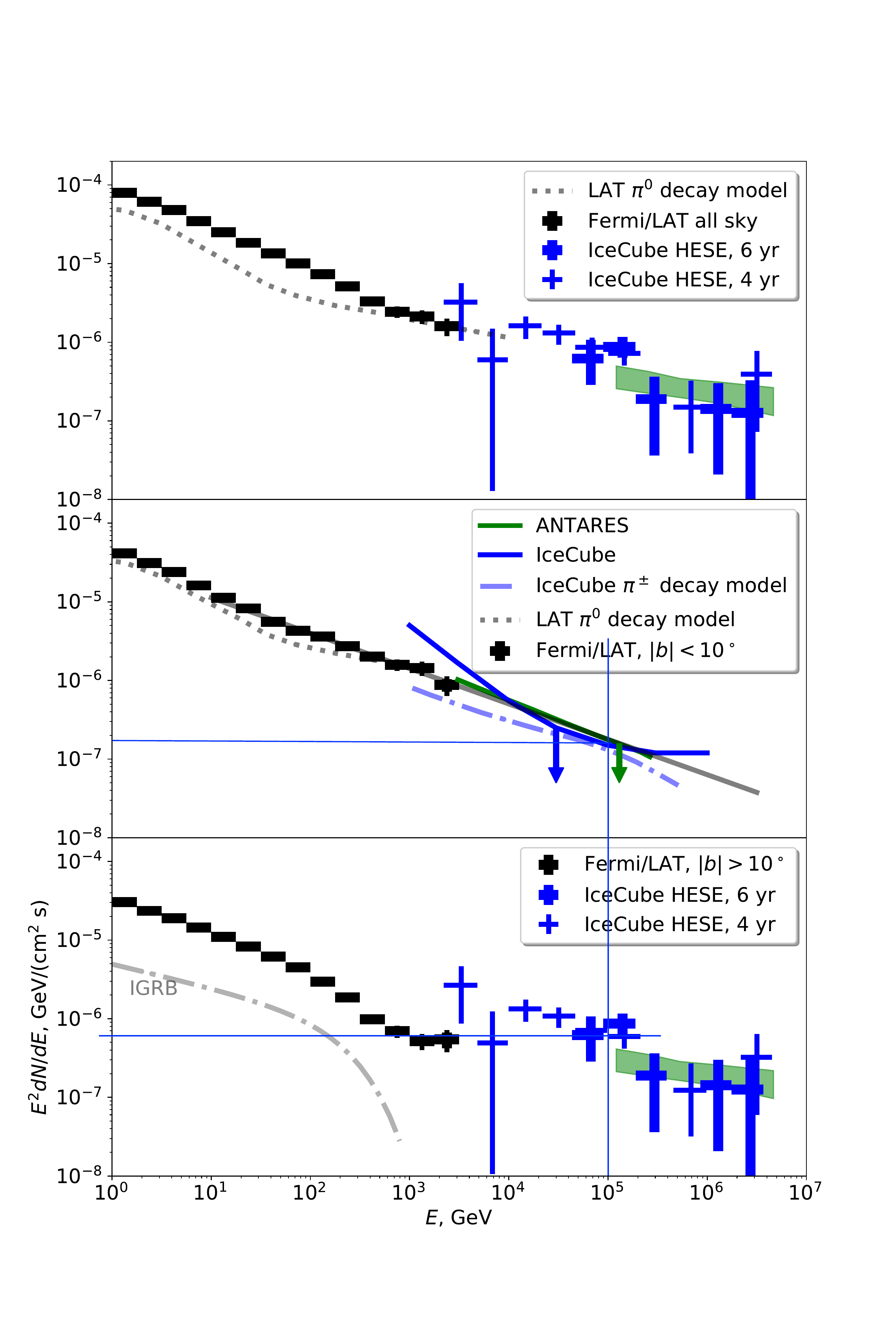}
\caption{
Top: the multi-messenger spectrum of the full sky. 
Fermi/LAT spectrum of $\gamma$-ray emission  is shown by black data points; thick and thin errorbars  show statistical and systematic uncertainties.  IceCube data are shown by blue data points  and by the green bow-tie (from Ref. \cite{icecube_icrc}). The dash-dotted curve shows the model of Galactic diffuse emission component from $\pi^0$ decays  \cite{diffuse_model}.  Middle: the Fermi/LAT spectrum of the Galactic Plane $|b|<10^\circ$  (black data points).  The blue dash-dotted curve shows model-dependent upper limit on neutrino flux derived under the assumption about particular shape of the $\pi^\pm$ decay spectrum \cite{galplane1}. Thin blue curve is an envelope of the upper bounds on the power-law spectra \cite{icecube_icrc} (see Appendix). The grey dotted curve shows the model of $\pi^0$ decay component of diffuse $\gamma$-ray flux  \cite{diffuse_model}.  Bottom: Fermi/LAT spectrum of $|b|>10^\circ$ region, compared to the IceCube neutrino flux measurements. The dash-dotted curve shows the best-fit model of the IGRB \cite{fermi_igrb}.}
\label{fig:spectrum}
\end{figure}

Figure \ref{fig:spectrum} compares Fermi/LAT $\gamma$-ray spectra of the full sky (upper panel), of the Galactic plane $|b|<10^\circ$ (middle panel) and at  Galactic latitudes $|b|>10^\circ$ (lower panel) with the neutrino spectra of the same sky regions  \cite{icecube_icrc,galplane,galplane1,antares_galplane}. In the spectra of the all-sky and the $|b|>10^\circ$ region we remove residual CR background, while the Galactic plane spectrum is calculated by subtracting high Galactic latitude background and residual cosmic ray contributions (see the Appendix for details). The $\gamma$-ray and neutrino all-sky flux and spectral slope measurements agree   in the overlapping multi-TeV band, confirming a previous analysis based on the high-energy extrapolation of the Fermi/LAT spectrum \cite{allsky}. Figure~1 also shows the model of diffuse $\gamma$-ray emission from pion decays derived from an all-sky analysis of the LAT data \cite{diffuse_model}. It is this component which is expected to have the neutrino counterpart, since pion decays produce simultaneously $\gamma$-rays ($\pi^0$ decays) and neutrinos ($\pi^\pm$ decays). The power-law extrapolation of the pion-decay model~\cite{diffuse_model} into the multi-TeV band agrees with the neutrino spectrum measured by IceCube. 

This agreement suggests the interpretation of the TeV $\gamma$-ray signal as the multi-messenger counterpart of the neutrino signal. However, the $\gamma$-ray flux below TeV is dominated by the emission from the Galactic plane, while only a moderate fraction of the neutrino flux in the 100\,TeV range comes from the Galactic plane \cite{galplane,galplane1,icecube_icrc,antares_galplane}. A consistent interpretation of the multi-TeV $\gamma$-ray flux should provide an explanation for this fact. If the multi-TeV $\gamma$-ray flux is the counterpart of the neutrino signal, the high Galactic latitude flux should have harder spectrum than the flux from the Galactic plane so that its relative contribution to the all-sky flux could grow with increasing energy. A hint of such a behavior can be noticed in the bottom panel of Fig. \ref{fig:spectrum} where a hardening of the flux is noticeable in the last two energy bins.

This hardening appears more pronounced in the analysis of the spectrum of the part of the sky at higher Galactic latitude, $|b|>20^\circ$, shown in Fig. \ref{fig:spectrum2}. In this figure we have removed contributions from resolved point sources, extragalactic isotropic diffuse $\gamma$-ray background (IGRB) and residual CR backgrounds thus leaving only the Galactic diffuse emission (see the Appendix).  The hardening of the  spectrum of diffuse emission  at  high Galactic latitudes starts at 300\,GeV and it can not be explained by instrumental effects (see the Appendix for details).  Below 300 GeV the spectrum is well fit by a smoothly broken power-law with the slope $\Gamma=2.906\pm 0.015$ in the 30--300\,GeV range.  The spectrum in the 0.3--3\,TeV range has the slope $\Gamma=2.09\pm 0.09$.

The most significant excess above the extrapolation of the  power-law valid below 300\,GeV is in the energy bin 1--1.7\,TeV. The model prediction of the number of photon counts in this bin is 16.4.   The observed number of counts is 39. The chance coincidence probability of such an excess is $1.5\times 10^{-6}$. In the energy bin 1.7--3.16\,TeV the expected number of counts is 3.8, while the observed one is 10. The chance coincidence probability of such an excess is $5.8\times 10^{-3}$. In the energy bin 0.3-1 TeV, the model prediction is $<66.5$ counts while the observed signal is 100 counts. The chance coincidence probability of the excess in this bin is  $8\times 10^{-5}$.  The energy-binning independent combined chance probability of the excess  above 300 GeV is less than $8\times 10^{-10}$.

\section{Interpretation}%
The multi-TeV band \gr\ flux at high and low Galactic latitude shown in Fig.~\ref{fig:spectrum} originates from the Milky Way. The low Galactic latitude flux is certainly dominated by the emission from decays of pions produced in interactions of Galactic CRs with interstellar matter in the Galactic disk. Since the \gr\ and neutrino fluxes from pion decays are comparable, the \gr\ flux measurement in the multi-TeV range can be used as an estimate for the minimal possible neutrino flux from the Galactic plane. One can see from the middle panel of Fig. \ref{fig:spectrum} that this lower bound on the neutrino flux is consistent with the upper limit derived by the IceCube and ANTARES telescopes \cite{galplane,galplane1,icecube_icrc,antares_galplane}. Combining the lower and upper limits one finds that the neutrino flux from the Galactic plane has to be  just at the level of the multi-TeV \gr\ flux from this part of the sky. 

More puzzling are the spectral characteristics of the multi-messenger signal at high Galactic latitudes.  The conventional high Galactic latitude diffuse emission components have soft spectra in the TeV range \cite{fermi_igrb} and can not explain the observed spectral hardening above 300 GeV.  The same is true for the IGRB, which is dominated by the cumulative flux of blazars \cite{fermi_igrb_blazars}, a special class of active galactic nuclei  which do not provide the dominant contribution to the neutrino signal \cite{blazars,blazars1}.  Thus, the observed hardening of the \gr\ spectrum has to be interpreted as due to the presence of a new Galactic $\gamma$-ray flux component above 300~GeV. It is this component which is the counterpart of the neutrino signal with comparable flux in the multi-TeV range. 

Only few source types could produce multi-TeV multi-messenger  emission on large angular scales at high Galactic latitude with a hard spectrum. One possibility is interactions of CRs forming a previously unknown component of the Galactic CR population.  If this new component would reside everywhere in the Galactic disk, an equivalent spectral hardening would be observed in the spectrum of  the Galactic plane---which is not the case. Instead, the hard spectrum CRs could either reside in our local Galactic environment, or be a part of a very large halo.

\begin{figure}
\vskip-0.7cm
\includegraphics[width=\linewidth]{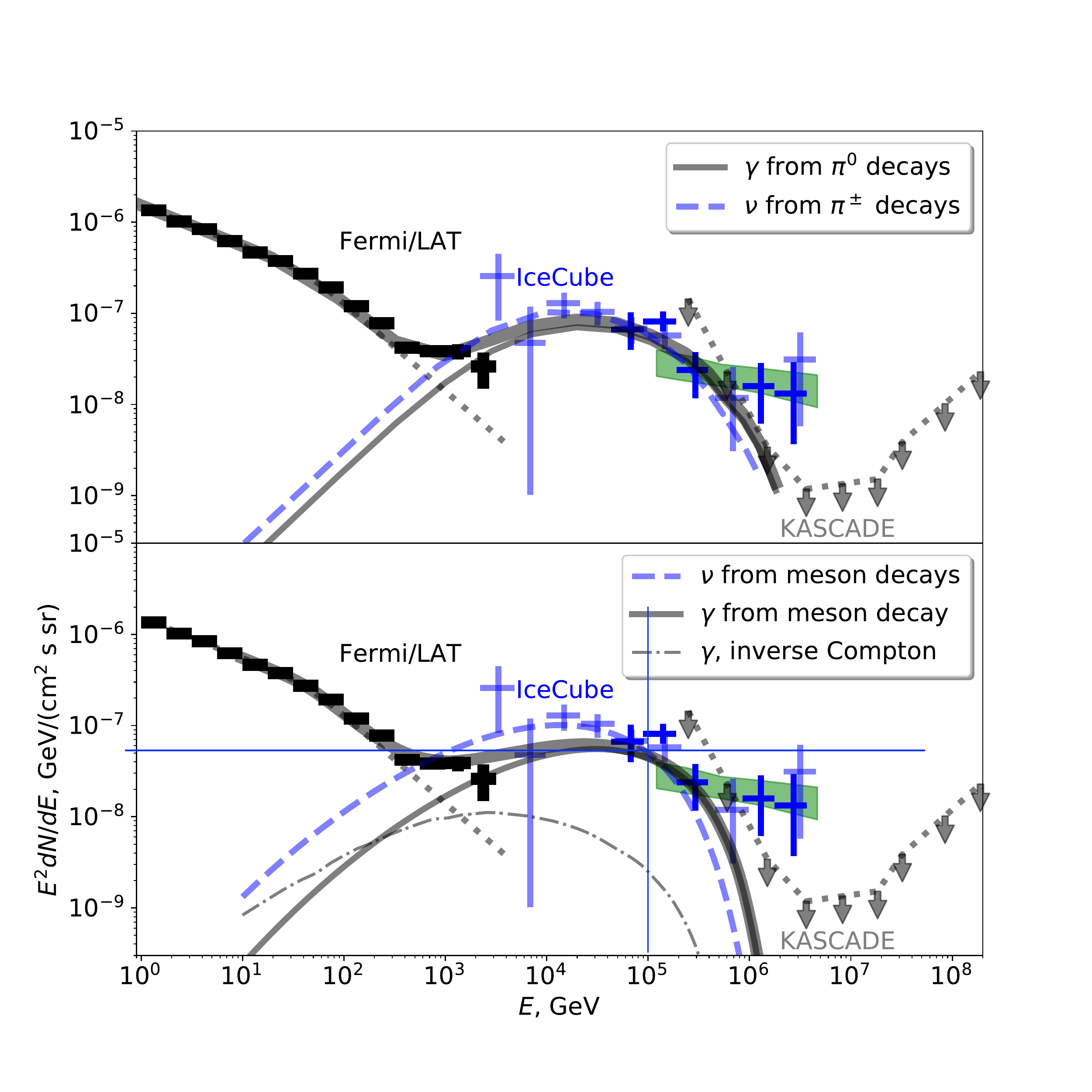}
\caption{High Galactic latitude emission for the local PeVatron (top) and DM (bottom) models. Thick and thin errorbars of Fermi/LAT data points (black) show statistical and systematic uncertainties, including the uncertainties of subtraction of IGRB and residual CR backgrounds.  Vertical arrows show KASCADE upper limits on the $\gamma$-ray flux from Northern sky  \cite{kascade}.  Solid thin  lines show the gamma-ray emission from the additional hard component.  Dashed lines show the neutrino emission. Dotted line shows a broken power-law fit to the sub-TeV $\gamma$-ray spectrum. Thick solid line shows the sum of the sub-TeV and additional hard $\gamma$-ray components.}
\label{fig:spectrum2}
\end{figure}

The local source of CRs with a hard spectrum reaching PeV energies (a "PeVatron") should be  a recent  and nearby  source, like e.g.\ the Vela supernova \cite{vela}. It should have injected CRs less than  $10^5$ year ago at a distance $d$ not larger than several hundred parsecs. These two conditions are required for the presence of PeV CRs which produce 10--100\,TeV neutrinos and the large angular extent $\Omega$ of the multi-messenger emission \cite{Andersen:2017yyg}. Cosmic rays with total energy  $U_{CR}\sim 10^{50}$\,erg injected by the PeVatron and loosing their energy on the time scale $t_{\rm pp}\simeq 1.5\times 10^8\left(n_{\rm ISM}/0.5\mbox{ cm}^{-3}\right)$\,yr in interactions with the interstellar medium of the density $n_{\rm ISM}\sim 0.5$~cm$^{-3}$ produce the $\gamma$-ray and neutrino flux $F=U_{CR}/(4\pi d^2 \Omega t_{pp})$ 
with magnitude
$$
F\sim
2\times 10^{-7}\left(\frac{\Omega}{2\pi{\rm sr}}\right)^{-1}
 \frac{n_{\rm ISM}}{0.5\rm/cm^{3}}
 \left( \frac{d}{0.3\rm kpc}\right)^{-2}
\frac{\rm GeV}{\rm cm^2\, s \,sr} .
$$
This flux estimate matches the observed signal level, cf.\ with Fig. 2. Otherwise, the high Galactic latitude emission could be from a very large (hundred kiloparsec) CR "storage" around the Milky Way disk \cite{halo_aharonian}. 

The local PeVatron model predicts strong  variability of the multi-messenger signal  across the sky. This variability is determined by the peculiarities of the energy-dependent spread of the CRs and  of the matter distribution in the local Galaxy. Low energy CRs which had no time to escape from the source region would not contribute to the large angular scale emission. This leads to a low-energy hardening of the spectrum, as shown in the top panel of Fig. 2 \cite{Andersen:2017yyg}.  In contrast, the signal is not expected to experience neither strong fluctuations nor a low-energy hardening in the large scale halo model~\cite{halo_aharonian}. 

An alternative possibility shown in the bottom panel of Fig. 2 is that decays of metastable DM particles $X$ with mass $m_X\simeq 5$~PeV generate photons and neutrinos~\cite{Berezinsky:1997hy,Feldstein:2013kka,Esmaili:2013gha}. The spectral shape of the decay mode $X\to \bar qq\to {\rm hadrons}$ is determined by Quantum Chromodynamics (QCD). Since at the end of the  QCD cascade quarks combine more easily to mesons than to baryons, mainly neutrinos and photons from pion decays are produced.  The $\gamma$-ray and neutrino flux measurements constrain the X particle lifetime  to be $\tau_X\sim 2\times 10^{27}\left(\Omega_X/\Omega_{DM}\right)^{-1}$~s, where $\Omega_X/\Omega_{DM}$ is the fraction of the DM in the form of $X$ particles \cite{Esmaili:2013gha,Feldstein:2013kka,murase}. Since the mass $m_X$  is above the unitarity limit~\cite{Griest:1989wd}, the $X$ particles were never in thermal equilibrium. They should have been  produced by gravitational interactions or other non-thermal processes  and may serve as a tool to study the earliest phases of the Universe. 

 The DM decay neutrino signal has a sizable extragalactic contribution, while its $\gamma$-ray  component in the TeV-PeV range has only the Galactic part. This leads to a systematically lower normalisation of the multi-TeV $\gamma$-ray component. The same is true for the large scale CR halo, which should be present around all galaxies, so that the neutrino flux is expected to have a significant extragalactic contribution. To the contrary, the neutrino and $\gamma$-ray components in the local PeVatron  model both originate from the Milky Way. The absence of the extragalactic component leads to similar $\gamma$-ray and neutrino fluxes   (see top panel of  Fig. 2). 
 
 The DM halo of the Galaxy is denser in the direction of the inner Galaxy. This means that in the  DM model, the flux from the inner Galaxy should be stronger than that from the outer Galaxy. However, the signal from the Galactic plane shown in Fig.~\ref{fig:spectrum} contains both the direction toward the Galactic center and the anticenter, from which the strongest and the weakest DM decay signal should be observed. We have verified that the expected excess of the DM decay signal from the Galactic Plane is consistent with the IceCube upper bounds on the Galactic plane flux. The fraction of the DM decay signal from the region $|b|<10^\circ$ is 0.22. Combining the information form Fig.~\ref{fig:spectrum} and Fig.~\ref{fig:spectrum2},  one can see that the neutrino flux from the high Galactic latitude region which is supposed to account for the full neutrino signal at high Galactic latitude at 100 TeV is at the level $6\times 10^{-7}$~GeV/cm$^2$s at this energy (cf. with the bottom panel of Fig.~\ref{fig:spectrum}). Re-scaling it by a factor $0.22/0.78\simeq 0.3$, one could check that the expected DM decay flux from the direction of the Galactic plane is at the level of $2\times 10^{-7}$~GeV/cm$^2$s, i.e.\ marginally consistent with the IceCube upper limit on the neutrino flux from the Galactic plane (the IceCube upper limit is exactly at the level of the flux estimate, which means that the signal of DM origin should soon reveal an excess toward the inner Galaxy). There is, however, one important reservation which should be added.  The IceCube upper limit on the Galactic emission is derived assuming certain spatial template for the signal distribution. This template does not correspond to the spatial template of the DM signal. Thus, the IceCube limit on the Galactic emission is not directly comparable to the DM model prediction. 
 
 For the local PeVatron model, there is no fixed spatial template because the source morphology is not known. No  excess toward the Galactic Plane is generically expected. In this respect, the IceCube limit on the Galactic emission component does not provide constraints on the local PeVatron model.

The distinction between possible models of the multi-messenger  signal based on  spectral or spatial characteristics will be possible with next generation instruments like the IceCube-Generation II \cite{gen2}, KM3NeT \cite{km3net} neutrino telescopes and the space-based $\gamma$-ray telescope  HERD\footnote{\protect\url{http://herd.ihep.ac.cn}} which will accumulate higher signal statistics. The detection of the $\gamma$-ray part of the signal by  ground-based telescopes like CTA\footnote{\protect\url{https://www.cta-observatory.org}}, LHAASO \cite{lhaaso} and CARPET~\cite{carpet} will be possible provided that a sufficiently high (by a factor $\sim 10^5$) rejection level of the CR background is achieved.    

\section{Conclusions}%
We have demonstrated that the properties of the large scale diffuse Galactic \gr\ flux in multi-TeV band are compatible with the flux and spectrum of the neutrino signal in 1-100 TeV range, so that the two signals may be considered as different components of one and the same "multi-messenger" signal in the multi-TeV sky.  The  $\gamma$-ray flux at high Galactic latitude exhibits a pronounced hardening above 300 GeV, while no hardening is observed in the low Galactic latitude flux. This effect explains the lower contribution from the Galactic plane to the neutrino signal at higher energies, as observed by IceCube.  We have suggested three possible models which could explain the observed hard spectrum high Galactic latitude multi-messenger emission above 300 GeV: (i) interactions of CRs injected by a recent nearby cosmic PeVatron, 
(ii) CR interactions in a large halo around the Milky Way, or 
(iii) decays of  DM particles. 

\acknowledgments
We would like to thank T.~Porter and V.~Savchenko for useful discussions of the data analysis.

\appendix

\section{Fermi LAT data selection and data analysis} 

Our analysis uses data of the Fermi/LAT telescope collected during the period between October 28, 2008 and December 15, 2017~\footnote{We discard the data between August 8 and October 28, 2008, as recommended by the Fermi Science Support Center for the analysis of the data above 30 GeV, see \protect\url{https://fermi.gsfc.nasa.gov/ssc/data/analysis/LAT\_caveats.html}}. We use  the {\tt ULTRACLEANVETO} class \cite{fermi_igrb,pass8} events which have the lowest residual CR contamination. 

The data are processed using the version v10r0p5 of the Fermi Science Tools\footnote{\protect\url{https://fermi.gsfc.nasa.gov/ssc/data/analysis/}}, via
a {\it gtselect} -- {\it gtmktime} -- {\it gtbin} -- {\it gtexposure} chain to produce the spectra of different parts of the sky using the "aperture photometry" approach~\footnote{\protect\url{https://fermi.gsfc.nasa.gov/ssc/data/analysis/scitools/aperture_photometry.html}}. The exposures for large regions of the sky are calculated averaging the exposures estimated on a grid of points with 10 degree spacing. 

\begin{figure}
\includegraphics[width=\linewidth]{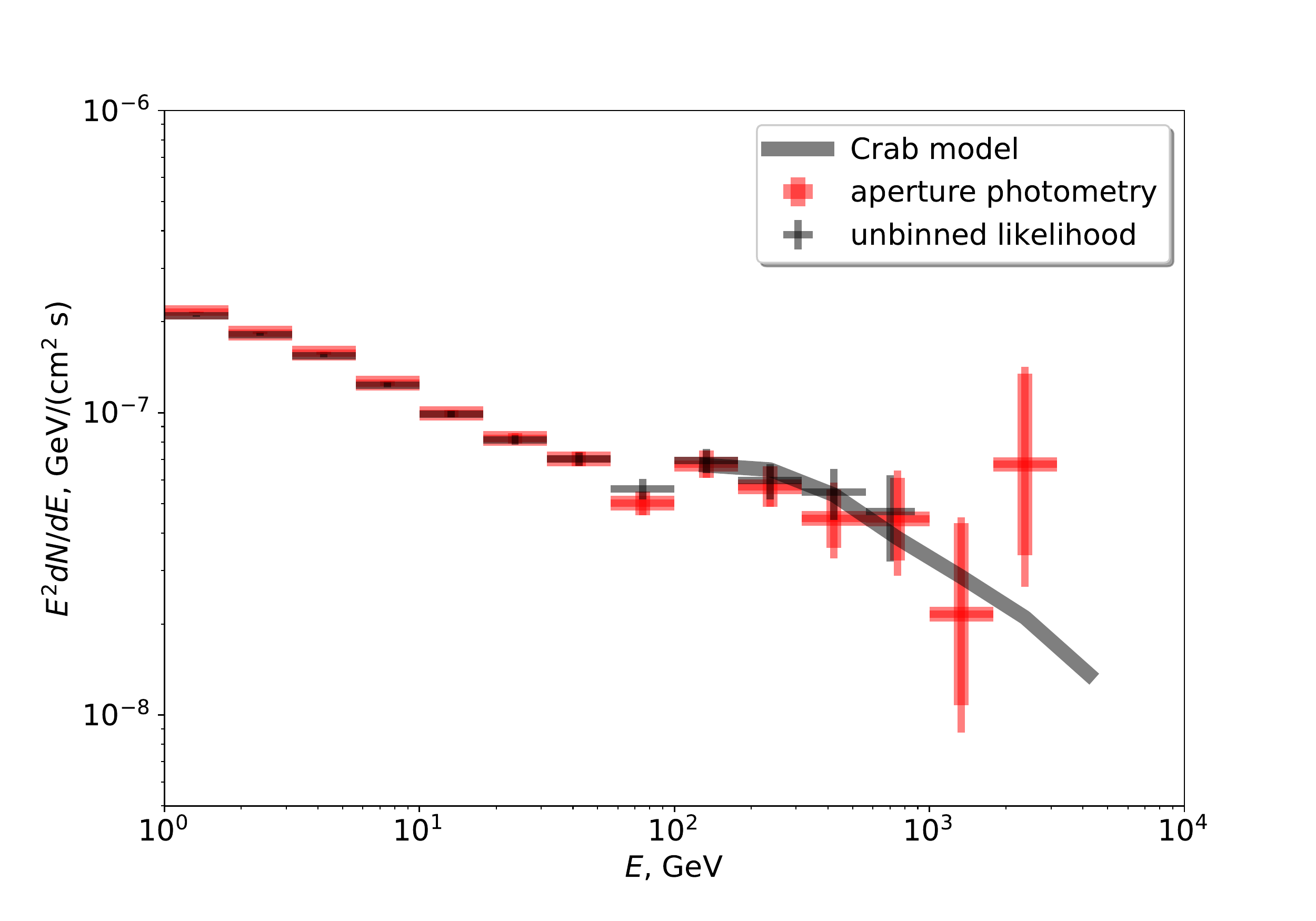}
\caption{Comparison of Fermi/LAT spectra of Crab extracted using unbinned likelihood (grey data points) and aperture photometry (red data points) methods. The grey thick line shows the Crab spectrum measured by ground-based $\gamma$-ray telescopes \cite{crab}. }
\label{fig:crab}
\end{figure}

We have verified that the spectra of isolated point sources extracted using this method are consistent with those extracted using an unbinned likelihood analysis~\footnote{\protect\url{https://fermi.gsfc.nasa.gov/ssc/data/analysis/scitools/likelihood_tutorial.html}}. As an example, Fig. \ref{fig:crab} shows a comparison of Crab spectra extracted using the two methods.  The two spectra are also extracted using two different photon selections: {\tt SOURCE} for the likelihood and {\tt ULTRACLEANVETO} for the aperture photometry. One can see that the  errorbars of the {\tt ULTRACLEANVETO} measurements (using aperture photometry) are somewhat larger because of the lower signal statistics. 

\subsection{LAT analysis in the multi-TeV band}

The spectral measurements based on the likelihood analysis do not extend into the TeV band because the public version of the {\it gtlike} tool has as analysis limit $\sim 850$ GeV. The Fermi Science Support Centre provides photon data in the energy range up to 10 TeV and information on the instrument characteristics (e.g. energy resolution, effective area) derived from Monte-Carlo simulations up to 3.16 TeV \footnote{\protect\url{http://www.slac.stanford.edu/exp/glast/groups/canda/lat_Performance.htm}}. These data could still be used for the aperture photometry analysis. 

The energy resolution of the telescope degrades in this energy band because of the increasing leakage of the signals produced by particle showers in the calorimeter and because of the saturation of the calorimeter crystals \cite{bruel12}. Still, reliable estimates of the energy are achieved up to at least 3 TeV, as described in Refs.  \cite{bruel12,pass8}. The energy resolution decreases from 10\% at 1 TeV to 25\% at 3 TeV.   In our analysis, we bin events in wide energy bins (4 bins per energy decade) which are much wider than the energy resolution over the entire analysis energy range. The most recent analysis results extend into the multi-TeV energy \cite{extended_galactic} thus validating the energy calibration in the TeV band based on the real data, via a direct comparison of the spectra derived from Fermi/LAT with the measurements by the ground-based $\gamma$-ray telescopes.  

The rapid degradation of the energy resolution might also result in the effect of a "pile-up" of higher energy events which were mis-reconstructed and attributed an energy close to a characteristic energy at which the energy resolution starts to worsen. In order to explore if pile-up effects might affect our spectral measurements, we have repeated the analysis using events from the {\tt EDISP2} and {\tt EDISP3} event sub-selections. This sub-selection is characterised by increasingly better energy reconstruction quality \footnote{\protect\url{https://fermi.gsfc.nasa.gov/ssc/data/analysis/documentation/Cicerone/Cicerone_LAT_IRFs/IRF_overview.html}}. Higher energy events with poor energy reconstruction which might produce a pile-up effect are naturally excluded in the high quality energy reconstruction event sub-samples. The spectra extracted using the {\tt EDISP2} and {\tt EDISP3} event sub-selections are consistent with those based on the full {\tt ULTRCLEANVETO} sample. 
 
Apart from the energy resolution, the systematic uncertainty of the effective area also keeps growing from 5\% at 100 GeV to $\ge 15$\% above 1 TeV \footnote{\protect\url{https://fermi.gsfc.nasa.gov/ssc/data/analysis/LAT_caveats.html}}. Extrapolating into the multi-TeV energy range, one finds that the uncertainty exceeds 25\%, i.e. the overall effective area is uncertain by a factor of 2.  Such a large uncertainty motivates a cross-calibration of the Fermi/LAT flux measurements with those of ground-based $\gamma$-ray telescopes, as  described below. 

Most of the observations of astronomical sources in the multi-TeV band are done using ground-based $\gamma$-ray telescopes, including Imaging Atmospheric Cherenkov Telescope (IACT) systems (HESS, MAGIC and VERITAS) and air shower arrays (MILAGRO, HAWC, ARGO-YBJ, Tibet-AS$\gamma$). The validation of the instrument response function of Fermi/LAT is possible via a cross-calibration of the LAT observations of selected sources with ground-based $\gamma$-ray telescope observations.   Taking into account the limited statistics of the LAT data in the multi-TeV band, we perform a comparison of spectral measurements of selected calibration sources for the stacked source signal rather than on a source-by-source basis. The sources selected for the cross-calibration purposes should have the following basic characteristics which make them suitable for the calibration analysis.

First, we require that the sources are steady in the TeV band. This excludes active galactic nuclei which are known to be strongly variable. Among the Galactic sources pulsar wind nebulae and supernova remnants are suitable.  All the pulsar wind nebulae and supernova remnants are extended sources. A further selection criterion for the calibration sources is the requirement that the source should have a well constrained spatial morphology. Uncertainties in the spatial structure of the source lead to uncertainties in the re-calculation of the flux measurements for different telescopes because of differing telescope point spread functions. The spatial morphology constraint leaves only a handful of isolated TeV $\gamma$-ray sources for the cross-calibration analysis. These sources are listed in Table \ref{tab:sources}.   

\begin{table*}
\begin{tabular}{llllll}
\hline
Name & Ra & DEC & Radius   &Counts & Spectral reference\\	     
\hline
Crab & 83.633 & 22.019 & $0.2^\circ$  &4 & \cite{crab}\\
Vela Jr & 133.2 & -46.5 &1.4$^\circ$ & 7 &\cite{velajr}\\
Vela X & 128.3 & -45.2 & 1.4$^\circ$ & 2 &\cite{velax}\\
RX J1713.7-3946 & 258.4 & -39.8 & $0.8^\circ$ & 2 &\cite{rxj1713}\\
HESS J1825-137 & 276.4 & -13.9 & $1.0^\circ$ & 3&\cite{hess1825}\\
\hline
Galactic plane $40^\circ<l<100^\circ$, $|b|<5^\circ$ &&&&19&\cite{argo,milagro}\\
\hline
Galactic plane $|b|<10^\circ$ &&&&209\\
\hline

All sky & & &&282\\
\hline
\end{tabular}
\caption{Count statistics of isolated sources on the Fermi/LAT sky map in the $E>1$~TeV energy range.}
\label{tab:sources}
\end{table*}

\begin{figure}
\includegraphics[width=\linewidth]{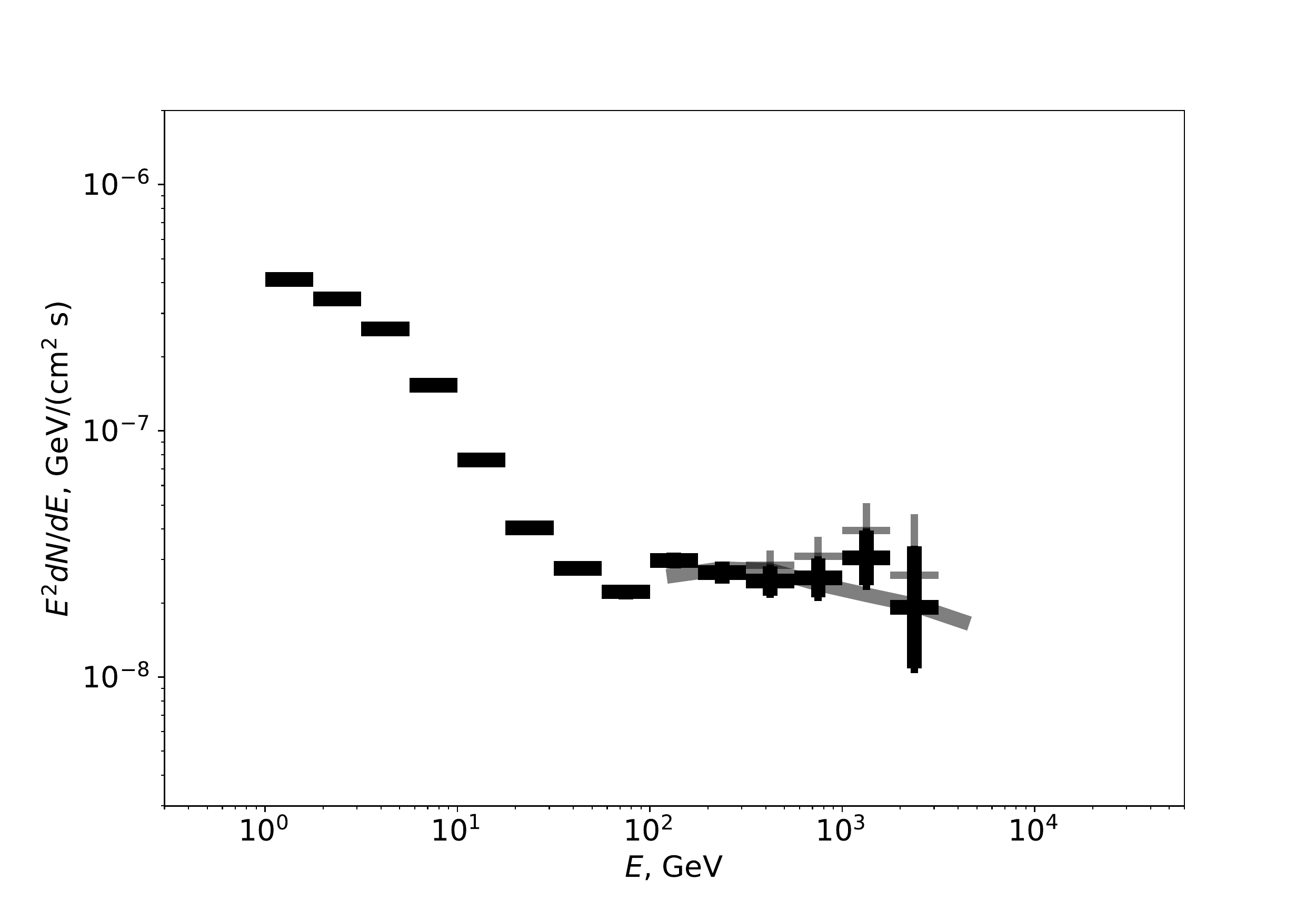}
\caption{ Stacked Fermi/LAT spectrum of isolated sources listed in Table~1. Grey thin data points show the spectrum extracted assuming effective area calculated with {\it gtexposure} tool, black thick data points show the spectrum calculated with effective area renormalised by the cross-calibration factor $\kappa$ (black data points). Grey thick line shows the average over the sources model spectrum.  }
\label{fig:stacked}
\end{figure}

The stacked spectrum of selected sources is shown in Fig. \ref{fig:stacked}. For each source, the source signal was extracted from a circle of the radius listed in the 4th column of Table \ref{tab:sources} (the circle radius is adjusted to cover the source extent and to include the wings of the LAT point spread function). The background is estimated from circles with radius either equal to the signal circle radius (for extended sources) or to $1^\circ$ (for Crab) and shifted from the source positions along constant Galactic latitude.  The reference spectral model is obtained by averaging the model spectra of individual sources derived in the References listed in Table \ref{tab:sources}. One can see that the Fermi/LAT measurement in the TeV band generally agrees with the spectral measurements done using ground-based $\gamma$-ray telescopes. The calibration of Fermi/LAT using ground-based measurements could be explicitly forced via a re-normalisation of the signal in the multi-TeV energy range on the model stacked source spectrum, as shown in Fig. \ref{fig:stacked}. The comparison of the LAT and ground-based telescope data in multi-TeV band shows that the agreement of the flux measurements for the stacked source spectrum is reached if the LAT flux is renormalised by $\simeq 20$\% (comparable to the systematic error \footnote{\protect\url{https://fermi.gsfc.nasa.gov/ssc/data/analysis/LAT_caveats.html}}).  We apply a cross-calibration factor $\kappa=1-c\log\left(E/100\mbox{ GeV}\right)$, with $c=0.25\pm 0.12$  at the energy $E>300$~GeV to achieve better consistence with the ground-based telescope measurements, following a practice common in X-ray data analysis, see e.g.\ \url{https://heasarc.gsfc.nasa.gov/docs/heasarc/caldb/caldb_xcal.html}. The uncertainty of the parameter of the cross-calibration factor  $c$ is taken into account as an additional systematic error. 

\begin{figure}
\includegraphics[width=\linewidth]{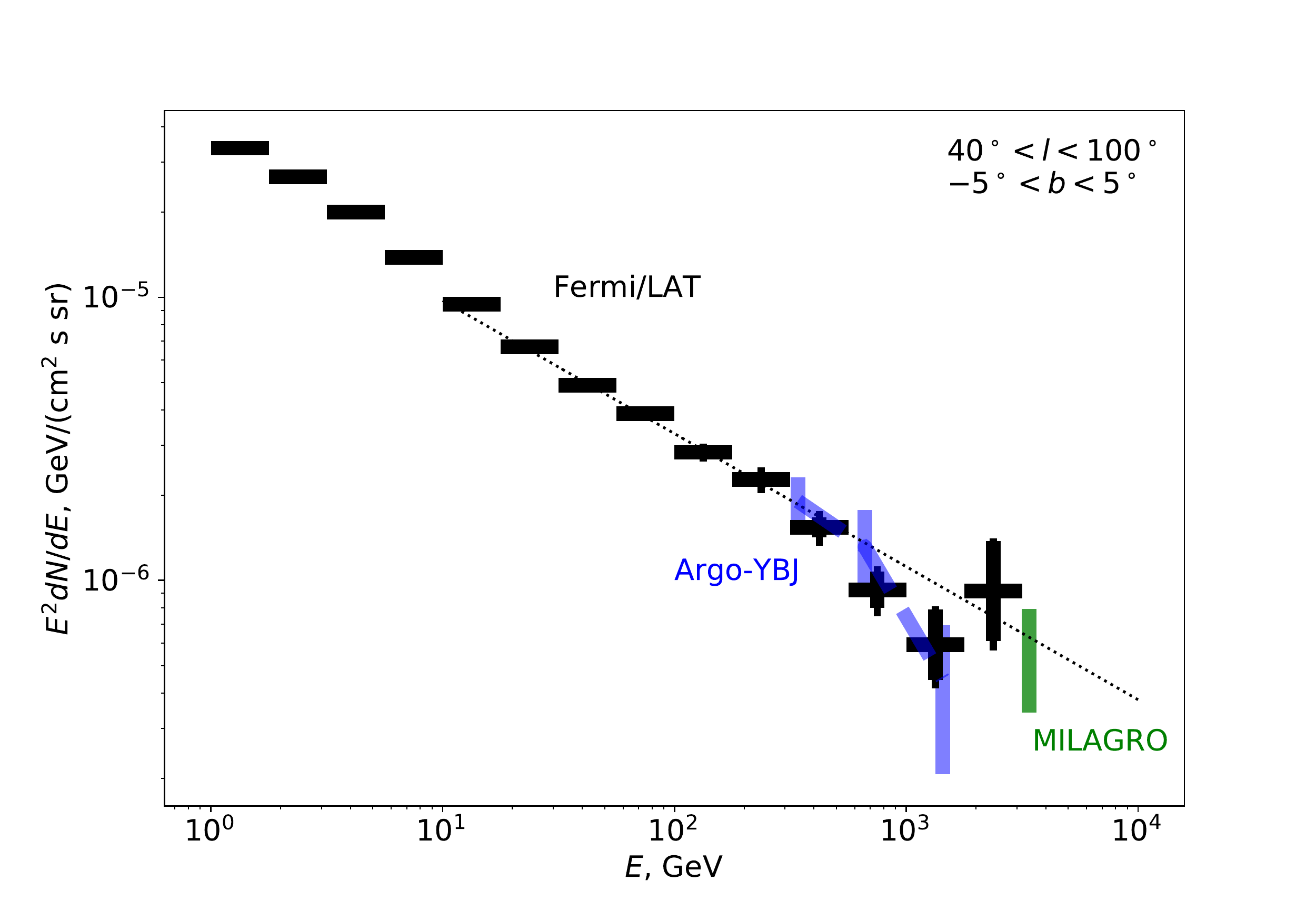}
\caption{Combined Fermi/LAT, ARGO-YBJ \cite{argo} and MILARGO \cite{milagro} spectrum of the $40^\circ<l<100^\circ$ stretch of the Galactic plane. Grey and black data points show the measurements (notations are the same as in Figs. 1,2 of the main text). Green data point is from MILAGRO \cite{milagro}, blue data points are from ARGO-YBJ \cite{argo}. \label{S3}}
\end{figure}

An additional cross-check of the Fermi/LAT calibration in the multi-TeV band can be extracted from a comparison of the spectra for the part of the Galactic plane observed  by  ARGO-YBJ \cite{argo} and MILAGRO \cite{milagro} air shower arrays. Figure~\ref{S3} shows the combined Fermi/LAT $+$ ARGO-YBJ $+$ MILAGRO spectrum of the $40^\circ<l<100^\circ$ region of the Galactic plane. The high Galactic latitude signal discussed in the main text is used as the background estimate for this region. One can see that the Fermi/LAT measurements agree with both ARGO-YBJ and MILAGRO data for this sky region. To calculate the Fermi/LAT spectrum of an extended region of the sky, we average the exposure  calculated on a grid of points with $10^\circ$ spacing using the {\it gtexposure} tool.  

\subsection{Residual CR background estimate and systematic effects}

\begin{figure} 
\includegraphics[width=\linewidth]{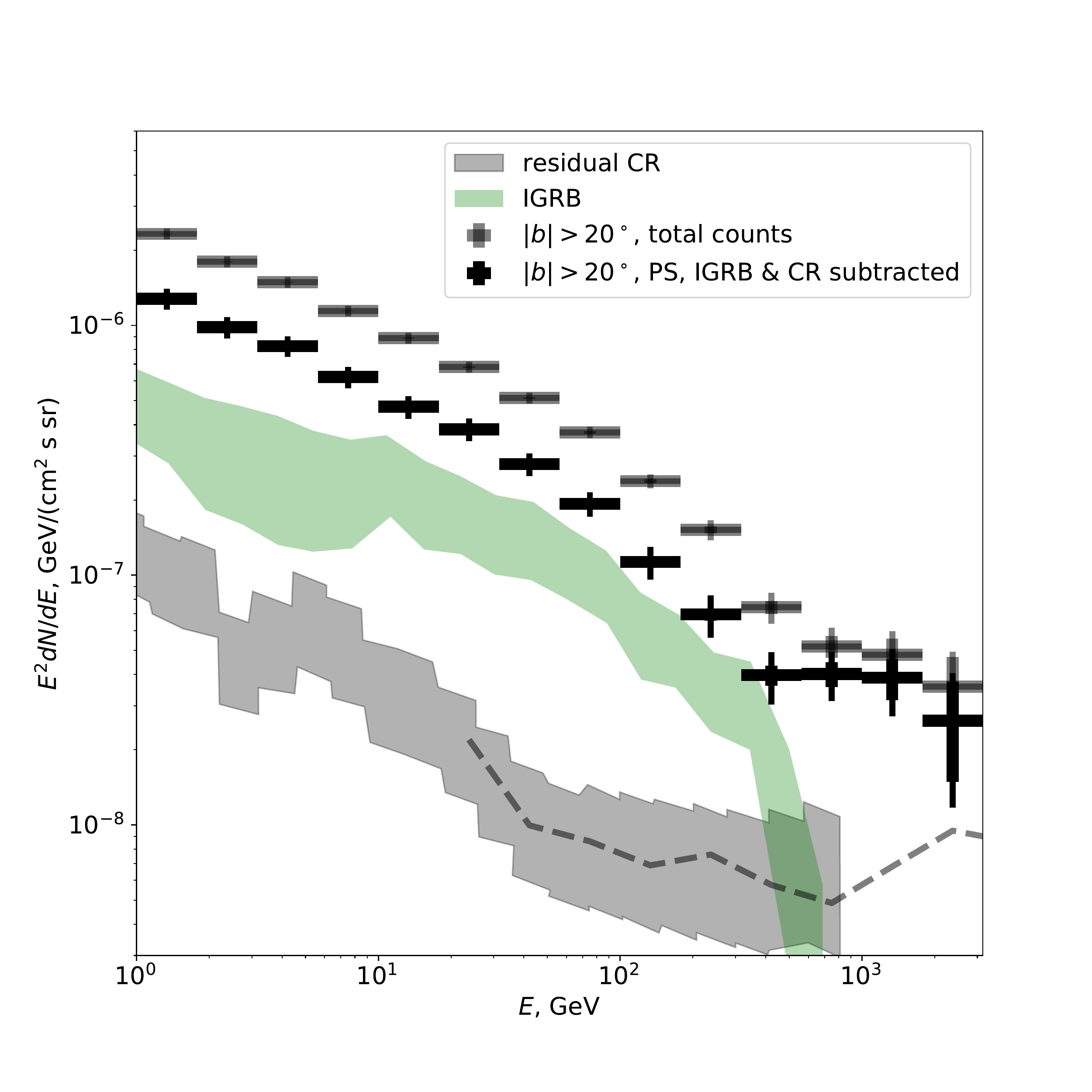}
\caption{ Fermi/LAT spectrum of diffuse $\gamma$-ray emission from high Galactic latitude (black data points and gray data point above 3 TeV) compared to the level of residual CR background (dotted thin line) derived in Ref. \cite{fermi_igrb} extended to the energy range above 1 TeV (thick dotted line). Red thin dotted line shows the increase of residual CR background under the assumption of hardening of the CR count rate power-law slope by 1. Red data points show the high Galactic latitude diffuse emission spectrum calculated assuming this higher residual CR background. Green shaded band shows the range of uncertainty of IGRB derived in Ref. \cite{fermi_igrb}. Grey data points below 3 TeV show total high Galactic latitude flux without subtraction of catalog sources, isotropic diffuse $\gamma$-ray background and residual CR background contributions.  }
\label{fig:cr}
\end{figure}

The {\tt ULTRACLEANVETO} event sample contains residual charged CR background events which are arriving from random directions on the sky and could mimic a nearly isotropic $\gamma$-ray signal. A study of the residual CR background contamination in the related sub-selection of ULTRACLEAN events with additional veto imposed to reduce CR background  was reported in Ref. \cite{fermi_igrb} for the PASS7 event selection. This study shows that the level of residual CR background at 850 GeV is at the level of 10\% of the $\gamma$-ray flux from the high Galactic latitude region in this energy range, as shown in Fig. \ref{fig:cr}. This study has derived the residual CR background count rate which follows a power-law dependence on equivalent $\gamma$-ray energy above 50~GeV. The recalculation of this power law into an equivalent diffuse emission flux is performed by dividing the residual CR count rate by the energy-dependent effective area. This results in the residual CR background flux shown by dashed line in Fig. \ref{fig:cr}. Applying the same method we extend the residual CR background flux model to the energy range above 1 TeV, assuming that the power law for the CR count rate extends with the same slope into the multi-TeV energy range. One can see that the residual CR flux could not provide the dominant contribution to the high Galactic latitude emission in the multi-TeV range.

\noindent

\begin{figure} 
\includegraphics[width=\linewidth]{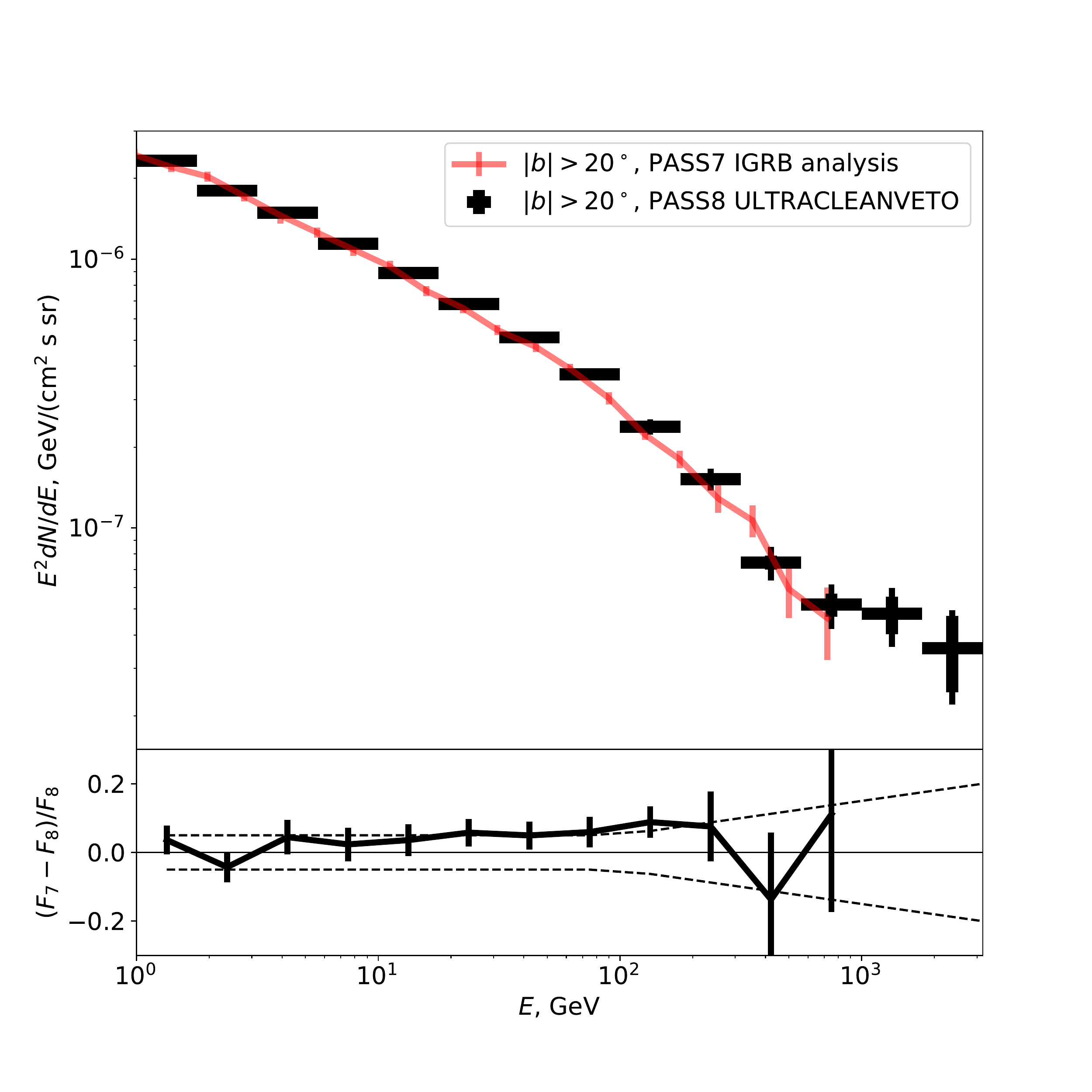}
\caption{Comparison of the total count ($\gamma$-ray $+$ residual cosmic ray background) spectra of the sky regions $|b|>20^\circ$ extracted using PASS7 IGRB and PASS8 ULTRACLEANVETO event selections. Top panel shows the spectral data, bottom panel shows the difference between the two spectral measurements compared to the systematic error delimited by the dashed line. PASS7 spectrum is from Ref. \cite{fermi_igrb}.}
\label{fig:pass7_pass8}
\end{figure}

The study of Ref. \cite{fermi_igrb} was based on the IGRB class of the PASS7 event selection. This class is  a sub-class of the ULTRACLEAN event class with additional veto conditions applied to reduce the residual CR background level. Our analysis is based on the ULTRACLEANVETO class of the PASS8 event selection. The IGRB class of PASS7 is not publicly available and a direct comparison of the ULTRACLEANVETO/PASS8  and IGRB/PASS7 event samples is not possible. However, we have verified that the total ($\gamma$-ray $+$ residual cosmic ray background) fluxes of the sky regions $|b|>20^\circ$ in the two event classes are compatible within the systematic uncertainty. This is shown in Fig. \ref{fig:pass7_pass8}.

\begin{figure} 
\includegraphics[width=\linewidth]{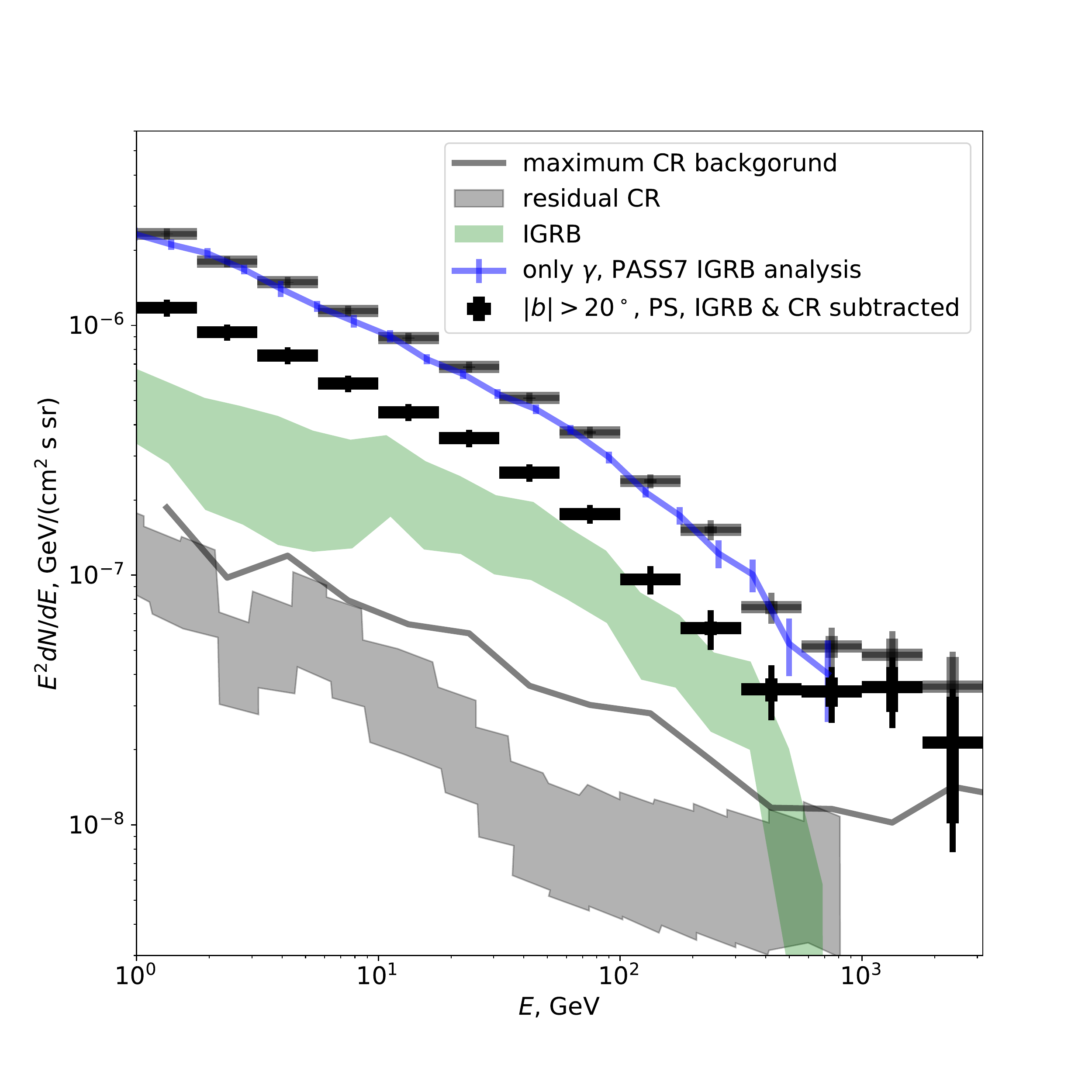}
\caption{ Same as in Fig. \ref{fig:cr} but for the "maximal possible" residual cosmic ray background estimate. Grey solid curve shows the maximal possible residual cosmic ray background in PASS8 ULTRACLEANVETO event selection.} 
\label{fig:cr_max}
\end{figure}

There is a systematic shift between the PASS7 and PASS8 measurements. It is possible, in principle, that this shift is due to higher level of residual CR background in the PASS8 event selection rather than to a different modelling of the instrument response functions. Adopting this hypothesis, one can estimate "conservatively" the maximal possible level of residual cosmic ray background in the PASS8 ULTRACLEANVETO event sample by subtracting the \gr\ flux of the $|b|>20^\circ$ part of the sky derived in Ref. \cite{fermi_igrb} from the total count spectrum calculated for the PASS8 ULTRACLEANVETO event selection. This estimate of the maximal possible residual cosmic ray background is shown in Fig. \ref{fig:cr_max}. One can see that subtracting the maximal possible residual CR background does not alter the properties of the hard excess above 300\,GeV.

\noindent

\subsection{Point source flux subtraction}

\begin{figure} 
\includegraphics[width=\linewidth]{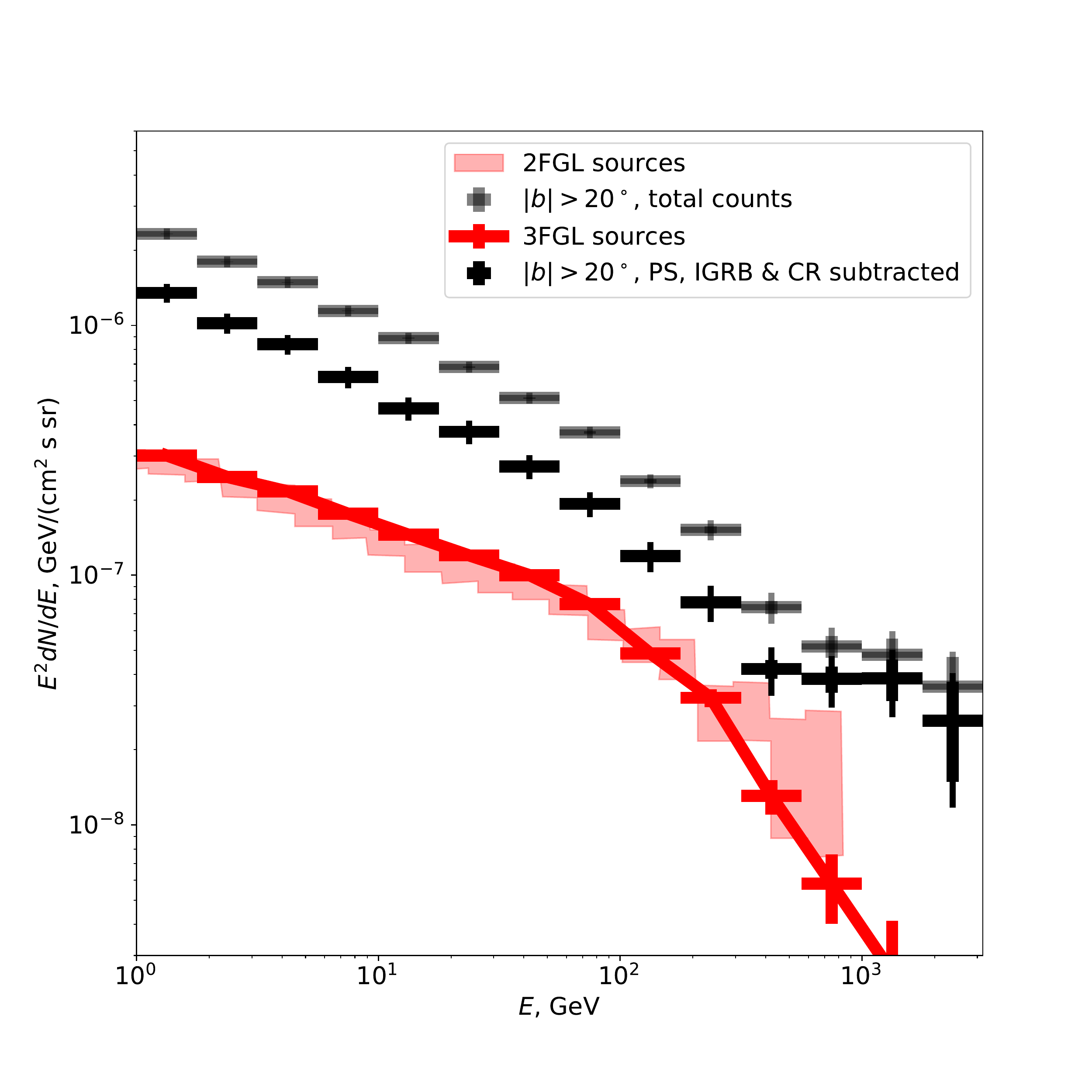}
\caption{ Point source contribution to the flux of  high Galactic latitude emission. Rose shaded range shows the estimate from Ref. \cite{fermi_igrb}. Red data points show the calculation based on 3FGL catalog \cite{fermi_cat}.} 
\label{fig:ps}
\end{figure}
\noindent

 Figure \ref{fig:spectrum2} of the main text shows the spectrum of high Galactic latitude $\gamma$-ray emission after subtraction of not only the residual CR and isotropic  diffuse $\gamma$-ray backgrounds, but also of the flux from isolated catalog sources \cite{fermi_cat}.

For the determination  of the point-source flux we have used the method described in Ref.~\cite{Neronov:2010vv}. First we stack the angular distributions of photons around the brightest point sources to obtain a measurement of the point-spread function in each energy interval. Next, we take all sources from the Fermi catalog \cite{fermi_cat} and perform a stacking analysis of the source and background signal around them. For the calculation of the background we exclude photons in circles with radius equal to the 95\% containment radius of the point-spread function and shuffle the remaining photons in Galactic longitude $l$ according to the Fermi exposure. 

The  point-source flux for $|b|>20^\circ$ in the 10 GeV band constitutes the same fraction of the total flux as found in Ref. \cite{fermi_igrb}. However, we find a smaller point-source fraction in the energy range above 100~GeV, see Fig. \ref{fig:ps}. The signal of point sources in this energy range is dominated by the contribution of BL Lacs and unidentified blazars~\cite{fermi_igrb_blazars}, most of which are also BL Lacs. The contribution of unknown types of sources is less then 10 \% at 100 GeV and reduces to zero at $E>300$ GeV. Taking into account that the point-spread function in this energy range has a narrow width, we have verified the point-source flux calculation by summing the photons within circles of 1 degree around the catalog source positions and estimating the remaining diffuse flux from the photon counts outside the source circles.

\section{IceCube upper limit on the Galactic Plane signal}

The upper limit on the Galactic Plane neutrino signal shown in the bottom panel of Fig. 1 is plotted as a envelope curve of the 90\% confidence level upper limits on the power-law type spectra as derived in Ref. \cite{icecube_icrc}. The envelope curve is tangent to the set of straight lines representing upper bounds on the flux for different values of the power-law slope.

\bibliography{references}

\begin{thebibliography}{44}
\expandafter\ifx\csname natexlab\endcsname\relax\def\natexlab#1{#1}\fi
\expandafter\ifx\csname bibnamefont\endcsname\relax
  \def\bibnamefont#1{#1}\fi
\expandafter\ifx\csname bibfnamefont\endcsname\relax
  \def\bibfnamefont#1{#1}\fi
\expandafter\ifx\csname citenamefont\endcsname\relax
  \def\citenamefont#1{#1}\fi
\expandafter\ifx\csname url\endcsname\relax
  \def\url#1{\texttt{#1}}\fi
\expandafter\ifx\csname urlprefix\endcsname\relax\def\urlprefix{URL }\fi
\providecommand{\bibinfo}[2]{#2}
\providecommand{\eprint}[2][]{\url{#2}}

\bibitem[{\citenamefont{{IceCube Collaboration}}(2013)}]{icecube_science}
\bibinfo{author}{\bibnamefont{{IceCube Collaboration}}},
  \bibinfo{journal}{Science} \textbf{\bibinfo{volume}{342}},
  \bibinfo{eid}{1242856} (\bibinfo{year}{2013}), \eprint{1311.5238}.

\bibitem[{\citenamefont{{Aartsen} et~al.}(2013)\citenamefont{{Aartsen},
  {Abbasi}, {Abdou}, {Ackermann}, {Adams}, {Aguilar}, {Ahlers}, {Altmann},
  {Auffenberg}, {Bai} et~al.}}]{icecube_pev}
\bibinfo{author}{\bibfnamefont{M.~G.} \bibnamefont{{Aartsen}}},
  \bibinfo{author}{\bibfnamefont{R.}~\bibnamefont{{Abbasi}}},
  \bibinfo{author}{\bibfnamefont{Y.}~\bibnamefont{{Abdou}}},
  \bibinfo{author}{\bibfnamefont{M.}~\bibnamefont{{Ackermann}}},
  \bibinfo{author}{\bibfnamefont{J.}~\bibnamefont{{Adams}}},
  \bibinfo{author}{\bibfnamefont{J.~A.} \bibnamefont{{Aguilar}}},
  \bibinfo{author}{\bibfnamefont{M.}~\bibnamefont{{Ahlers}}},
  \bibinfo{author}{\bibfnamefont{D.}~\bibnamefont{{Altmann}}},
  \bibinfo{author}{\bibfnamefont{J.}~\bibnamefont{{Auffenberg}}},
  \bibinfo{author}{\bibfnamefont{X.}~\bibnamefont{{Bai}}},
  \bibnamefont{et~al.}, \bibinfo{journal}{Physical Review Letters}
  \textbf{\bibinfo{volume}{111}}, \bibinfo{eid}{021103} (\bibinfo{year}{2013}),
  \eprint{1304.5356}.

\bibitem[{\citenamefont{{Halzen} and {Hooper}}(2002)}]{review1}
\bibinfo{author}{\bibfnamefont{F.}~\bibnamefont{{Halzen}}} \bibnamefont{and}
  \bibinfo{author}{\bibfnamefont{D.}~\bibnamefont{{Hooper}}},
  \bibinfo{journal}{Reports on Progress in Physics}
  \textbf{\bibinfo{volume}{65}}, \bibinfo{pages}{1025} (\bibinfo{year}{2002}),
  \eprint{astro-ph/0204527}.

\bibitem[{\citenamefont{Berezinsky}(1992)}]{Berezinsky:1991aa}
\bibinfo{author}{\bibfnamefont{V.~S.} \bibnamefont{Berezinsky}},
  \bibinfo{journal}{Nucl. Phys.} \textbf{\bibinfo{volume}{B380}},
  \bibinfo{pages}{478} (\bibinfo{year}{1992}).

\bibitem[{\citenamefont{Gondolo et~al.}(1993)\citenamefont{Gondolo, Gelmini,
  and Sarkar}}]{Gondolo:1991rn}
\bibinfo{author}{\bibfnamefont{P.}~\bibnamefont{Gondolo}},
  \bibinfo{author}{\bibfnamefont{G.}~\bibnamefont{Gelmini}}, \bibnamefont{and}
  \bibinfo{author}{\bibfnamefont{S.}~\bibnamefont{Sarkar}},
  \bibinfo{journal}{Nucl. Phys.} \textbf{\bibinfo{volume}{B392}},
  \bibinfo{pages}{111} (\bibinfo{year}{1993}), \eprint{hep-ph/9209236}.

\bibitem[{\citenamefont{{Ahlers} and {Halzen}}(2017)}]{review2}
\bibinfo{author}{\bibfnamefont{M.}~\bibnamefont{{Ahlers}}} \bibnamefont{and}
  \bibinfo{author}{\bibfnamefont{F.}~\bibnamefont{{Halzen}}},
  \bibinfo{journal}{Progress of Theoretical and Experimental Physics}
  \textbf{\bibinfo{volume}{2017}}, \bibinfo{eid}{12A105}
  (\bibinfo{year}{2017}).

\bibitem[{\citenamefont{{Franceschini}
  et~al.}(2008)\citenamefont{{Franceschini}, {Rodighiero}, and
  {Vaccari}}}]{franceschini}
\bibinfo{author}{\bibfnamefont{A.}~\bibnamefont{{Franceschini}}},
  \bibinfo{author}{\bibfnamefont{G.}~\bibnamefont{{Rodighiero}}},
  \bibnamefont{and}
  \bibinfo{author}{\bibfnamefont{M.}~\bibnamefont{{Vaccari}}},
  \bibinfo{journal}{A\&A} \textbf{\bibinfo{volume}{487}}, \bibinfo{pages}{837}
  (\bibinfo{year}{2008}), \eprint{0805.1841}.

\bibitem[{\citenamefont{Aharonian et~al.}(2008)\citenamefont{Aharonian,
  Buckley, Kifune, and Sinnis}}]{gound-based}
\bibinfo{author}{\bibfnamefont{F.}~\bibnamefont{Aharonian}},
  \bibinfo{author}{\bibfnamefont{J.}~\bibnamefont{Buckley}},
  \bibinfo{author}{\bibfnamefont{T.}~\bibnamefont{Kifune}}, \bibnamefont{and}
  \bibinfo{author}{\bibfnamefont{G.}~\bibnamefont{Sinnis}},
  \bibinfo{journal}{Reports on Progress in Physics}
  \textbf{\bibinfo{volume}{71}}, \bibinfo{pages}{096901}
  (\bibinfo{year}{2008}),
  \urlprefix\url{http://stacks.iop.org/0034-4885/71/i=9/a=096901}.

\bibitem[{\citenamefont{{Atwood} et~al.}(2009)\citenamefont{{Atwood}, {Abdo},
  {Ackermann}, {Althouse}, {Anderson}, {Axelsson}, {Baldini}, {Ballet}, {Band},
  {Barbiellini} et~al.}}]{atwood09}
\bibinfo{author}{\bibfnamefont{W.~B.} \bibnamefont{{Atwood}}},
  \bibinfo{author}{\bibfnamefont{A.~A.} \bibnamefont{{Abdo}}},
  \bibinfo{author}{\bibfnamefont{M.}~\bibnamefont{{Ackermann}}},
  \bibinfo{author}{\bibfnamefont{W.}~\bibnamefont{{Althouse}}},
  \bibinfo{author}{\bibfnamefont{B.}~\bibnamefont{{Anderson}}},
  \bibinfo{author}{\bibfnamefont{M.}~\bibnamefont{{Axelsson}}},
  \bibinfo{author}{\bibfnamefont{L.}~\bibnamefont{{Baldini}}},
  \bibinfo{author}{\bibfnamefont{J.}~\bibnamefont{{Ballet}}},
  \bibinfo{author}{\bibfnamefont{D.~L.} \bibnamefont{{Band}}},
  \bibinfo{author}{\bibfnamefont{G.}~\bibnamefont{{Barbiellini}}},
  \bibnamefont{et~al.}, \bibinfo{journal}{Ap.J.}
  \textbf{\bibinfo{volume}{697}}, \bibinfo{pages}{1071} (\bibinfo{year}{2009}),
  \eprint{0902.1089}.

\bibitem[{\citenamefont{{Bruel} and {Fermi-LAT Collaboration}}(2012)}]{bruel12}
\bibinfo{author}{\bibfnamefont{P.}~\bibnamefont{{Bruel}}} \bibnamefont{and}
  \bibinfo{author}{\bibnamefont{{Fermi-LAT Collaboration}}}, in
  \emph{\bibinfo{booktitle}{Journal of Physics Conference Series}}
  (\bibinfo{year}{2012}), vol. \bibinfo{volume}{404}, p.
  \bibinfo{pages}{012033}, \eprint{1210.2558}.

\bibitem[{\citenamefont{{Atwood} et~al.}(2013)\citenamefont{{Atwood}, {Albert},
  {Baldini}, {Tinivella}, {Bregeon}, {Pesce-Rollins}, {Sgr{\`o}}, {Bruel},
  {Charles}, {Drlica-Wagner} et~al.}}]{pass8}
\bibinfo{author}{\bibfnamefont{W.}~\bibnamefont{{Atwood}}},
  \bibinfo{author}{\bibfnamefont{A.}~\bibnamefont{{Albert}}},
  \bibinfo{author}{\bibfnamefont{L.}~\bibnamefont{{Baldini}}},
  \bibinfo{author}{\bibfnamefont{M.}~\bibnamefont{{Tinivella}}},
  \bibinfo{author}{\bibfnamefont{J.}~\bibnamefont{{Bregeon}}},
  \bibinfo{author}{\bibfnamefont{M.}~\bibnamefont{{Pesce-Rollins}}},
  \bibinfo{author}{\bibfnamefont{C.}~\bibnamefont{{Sgr{\`o}}}},
  \bibinfo{author}{\bibfnamefont{P.}~\bibnamefont{{Bruel}}},
  \bibinfo{author}{\bibfnamefont{E.}~\bibnamefont{{Charles}}},
  \bibinfo{author}{\bibfnamefont{A.}~\bibnamefont{{Drlica-Wagner}}},
  \bibnamefont{et~al.}, \bibinfo{journal}{arXiv:}
  \textbf{\bibinfo{volume}{1303.3514}} (\bibinfo{year}{2013}),
  \eprint{1303.3514}.

\bibitem[{\citenamefont{{Bartoli} et~al.}(2015)\citenamefont{{Bartoli},
  {Bernardini}, {Bi}, {Branchini}, {Budano}, {Camarri}, {Cao}, {Cardarelli},
  {Catalanotti}, {Chen} et~al.}}]{argo}
\bibinfo{author}{\bibfnamefont{B.}~\bibnamefont{{Bartoli}}},
  \bibinfo{author}{\bibfnamefont{P.}~\bibnamefont{{Bernardini}}},
  \bibinfo{author}{\bibfnamefont{X.~J.} \bibnamefont{{Bi}}},
  \bibinfo{author}{\bibfnamefont{P.}~\bibnamefont{{Branchini}}},
  \bibinfo{author}{\bibfnamefont{A.}~\bibnamefont{{Budano}}},
  \bibinfo{author}{\bibfnamefont{P.}~\bibnamefont{{Camarri}}},
  \bibinfo{author}{\bibfnamefont{Z.}~\bibnamefont{{Cao}}},
  \bibinfo{author}{\bibfnamefont{R.}~\bibnamefont{{Cardarelli}}},
  \bibinfo{author}{\bibfnamefont{S.}~\bibnamefont{{Catalanotti}}},
  \bibinfo{author}{\bibfnamefont{S.~Z.} \bibnamefont{{Chen}}},
  \bibnamefont{et~al.}, \bibinfo{journal}{Ap.J.}
  \textbf{\bibinfo{volume}{806}}, \bibinfo{eid}{20} (\bibinfo{year}{2015}),
  \eprint{1507.06758}.

\bibitem[{\citenamefont{{Atkins} et~al.}(2005)\citenamefont{{Atkins}, {Benbow},
  {Berley}, {Blaufuss}, {Coyne}, {De Young}, {Dingus}, {Dorfan}, {Ellsworth},
  {Fleysher} et~al.}}]{milagro}
\bibinfo{author}{\bibfnamefont{R.}~\bibnamefont{{Atkins}}},
  \bibinfo{author}{\bibfnamefont{W.}~\bibnamefont{{Benbow}}},
  \bibinfo{author}{\bibfnamefont{D.}~\bibnamefont{{Berley}}},
  \bibinfo{author}{\bibfnamefont{E.}~\bibnamefont{{Blaufuss}}},
  \bibinfo{author}{\bibfnamefont{D.~G.} \bibnamefont{{Coyne}}},
  \bibinfo{author}{\bibfnamefont{T.}~\bibnamefont{{De Young}}},
  \bibinfo{author}{\bibfnamefont{B.~L.} \bibnamefont{{Dingus}}},
  \bibinfo{author}{\bibfnamefont{D.~E.} \bibnamefont{{Dorfan}}},
  \bibinfo{author}{\bibfnamefont{R.~W.} \bibnamefont{{Ellsworth}}},
  \bibinfo{author}{\bibfnamefont{L.}~\bibnamefont{{Fleysher}}},
  \bibnamefont{et~al.}, \bibinfo{journal}{Physical Review Letters}
  \textbf{\bibinfo{volume}{95}}, \bibinfo{eid}{251103} (\bibinfo{year}{2005}),
  \eprint{astro-ph/0502303}.

\bibitem[{\citenamefont{{IceCube Collaboration}}(2017)}]{icecube_icrc}
\bibinfo{author}{\bibnamefont{{IceCube Collaboration}}},
  \bibinfo{journal}{Proc. of 35th International Cosmic Ray Conference, arXiv:}
  \textbf{\bibinfo{volume}{1710.01179}}, \bibinfo{pages}{981}
  (\bibinfo{year}{2017}).

\bibitem[{\citenamefont{{Neronov} and
  {Semikoz}}(2016{\natexlab{a}})}]{galplane}
\bibinfo{author}{\bibfnamefont{A.}~\bibnamefont{{Neronov}}} \bibnamefont{and}
  \bibinfo{author}{\bibfnamefont{D.}~\bibnamefont{{Semikoz}}},
  \bibinfo{journal}{Astroparticle Physics} \textbf{\bibinfo{volume}{75}},
  \bibinfo{pages}{60} (\bibinfo{year}{2016}{\natexlab{a}}),
  \eprint{1509.03522}.

\bibitem[{\citenamefont{{Aartsen}
  et~al.}(2017{\natexlab{a}})\citenamefont{{Aartsen}, {Ackermann}, {Adams},
  {Aguilar}, {Ahlers}, {Ahrens}, {Samarai}, {Altmann}, {Andeen}, {Anderson}
  et~al.}}]{galplane1}
\bibinfo{author}{\bibfnamefont{M.~G.} \bibnamefont{{Aartsen}}},
  \bibinfo{author}{\bibfnamefont{M.}~\bibnamefont{{Ackermann}}},
  \bibinfo{author}{\bibfnamefont{J.}~\bibnamefont{{Adams}}},
  \bibinfo{author}{\bibfnamefont{J.~A.} \bibnamefont{{Aguilar}}},
  \bibinfo{author}{\bibfnamefont{M.}~\bibnamefont{{Ahlers}}},
  \bibinfo{author}{\bibfnamefont{M.}~\bibnamefont{{Ahrens}}},
  \bibinfo{author}{\bibfnamefont{I.~A.} \bibnamefont{{Samarai}}},
  \bibinfo{author}{\bibfnamefont{D.}~\bibnamefont{{Altmann}}},
  \bibinfo{author}{\bibfnamefont{K.}~\bibnamefont{{Andeen}}},
  \bibinfo{author}{\bibfnamefont{T.}~\bibnamefont{{Anderson}}},
  \bibnamefont{et~al.}, \bibinfo{journal}{Ap.J.}
  \textbf{\bibinfo{volume}{849}}, \bibinfo{eid}{67}
  (\bibinfo{year}{2017}{\natexlab{a}}).

\bibitem[{\citenamefont{{Albert} et~al.}(2017)\citenamefont{{Albert},
  {Andr{\'e}}, {Anghinolfi}, {Anton}, {Ardid}, {Aubert}, {Avgitas}, {Baret},
  {Barrios-Mart{\'{\i}}}, {Basa} et~al.}}]{antares_galplane}
\bibinfo{author}{\bibfnamefont{A.}~\bibnamefont{{Albert}}},
  \bibinfo{author}{\bibfnamefont{M.}~\bibnamefont{{Andr{\'e}}}},
  \bibinfo{author}{\bibfnamefont{M.}~\bibnamefont{{Anghinolfi}}},
  \bibinfo{author}{\bibfnamefont{G.}~\bibnamefont{{Anton}}},
  \bibinfo{author}{\bibfnamefont{M.}~\bibnamefont{{Ardid}}},
  \bibinfo{author}{\bibfnamefont{J.-J.} \bibnamefont{{Aubert}}},
  \bibinfo{author}{\bibfnamefont{T.}~\bibnamefont{{Avgitas}}},
  \bibinfo{author}{\bibfnamefont{B.}~\bibnamefont{{Baret}}},
  \bibinfo{author}{\bibfnamefont{J.}~\bibnamefont{{Barrios-Mart{\'{\i}}}}},
  \bibinfo{author}{\bibfnamefont{S.}~\bibnamefont{{Basa}}},
  \bibnamefont{et~al.}, \bibinfo{journal}{Phys. Rev. D}
  \textbf{\bibinfo{volume}{96}}, \bibinfo{eid}{062001} (\bibinfo{year}{2017}),
  \eprint{1705.00497}.

\bibitem[{\citenamefont{{Neronov} and {Semikoz}}(2016{\natexlab{b}})}]{allsky}
\bibinfo{author}{\bibfnamefont{A.}~\bibnamefont{{Neronov}}} \bibnamefont{and}
  \bibinfo{author}{\bibfnamefont{D.}~\bibnamefont{{Semikoz}}},
  \bibinfo{journal}{Astroparticle Physics} \textbf{\bibinfo{volume}{72}},
  \bibinfo{pages}{32} (\bibinfo{year}{2016}{\natexlab{b}}), \eprint{1412.1690}.

\bibitem[{\citenamefont{{Acero} et~al.}(2016)\citenamefont{{Acero},
  {Ackermann}, {Ajello}, {Albert}, {Baldini}, {Ballet}, {Barbiellini},
  {Bastieri}, {Bellazzini}, {Bissaldi} et~al.}}]{diffuse_model}
\bibinfo{author}{\bibfnamefont{F.}~\bibnamefont{{Acero}}},
  \bibinfo{author}{\bibfnamefont{M.}~\bibnamefont{{Ackermann}}},
  \bibinfo{author}{\bibfnamefont{M.}~\bibnamefont{{Ajello}}},
  \bibinfo{author}{\bibfnamefont{A.}~\bibnamefont{{Albert}}},
  \bibinfo{author}{\bibfnamefont{L.}~\bibnamefont{{Baldini}}},
  \bibinfo{author}{\bibfnamefont{J.}~\bibnamefont{{Ballet}}},
  \bibinfo{author}{\bibfnamefont{G.}~\bibnamefont{{Barbiellini}}},
  \bibinfo{author}{\bibfnamefont{D.}~\bibnamefont{{Bastieri}}},
  \bibinfo{author}{\bibfnamefont{R.}~\bibnamefont{{Bellazzini}}},
  \bibinfo{author}{\bibfnamefont{E.}~\bibnamefont{{Bissaldi}}},
  \bibnamefont{et~al.}, \bibinfo{journal}{Ap.J.Supp.}
  \textbf{\bibinfo{volume}{223}}, \bibinfo{eid}{26} (\bibinfo{year}{2016}),
  \eprint{1602.07246}.

\bibitem[{\citenamefont{{Ackermann} et~al.}(2015)\citenamefont{{Ackermann},
  {Ajello}, {Albert}, {Atwood}, {Baldini}, {Ballet}, {Barbiellini}, {Bastieri},
  {Bechtol}, {Bellazzini} et~al.}}]{fermi_igrb}
\bibinfo{author}{\bibfnamefont{M.}~\bibnamefont{{Ackermann}}},
  \bibinfo{author}{\bibfnamefont{M.}~\bibnamefont{{Ajello}}},
  \bibinfo{author}{\bibfnamefont{A.}~\bibnamefont{{Albert}}},
  \bibinfo{author}{\bibfnamefont{W.~B.} \bibnamefont{{Atwood}}},
  \bibinfo{author}{\bibfnamefont{L.}~\bibnamefont{{Baldini}}},
  \bibinfo{author}{\bibfnamefont{J.}~\bibnamefont{{Ballet}}},
  \bibinfo{author}{\bibfnamefont{G.}~\bibnamefont{{Barbiellini}}},
  \bibinfo{author}{\bibfnamefont{D.}~\bibnamefont{{Bastieri}}},
  \bibinfo{author}{\bibfnamefont{K.}~\bibnamefont{{Bechtol}}},
  \bibinfo{author}{\bibfnamefont{R.}~\bibnamefont{{Bellazzini}}},
  \bibnamefont{et~al.}, \bibinfo{journal}{Ap.J.}
  \textbf{\bibinfo{volume}{799}}, \bibinfo{eid}{86} (\bibinfo{year}{2015}),
  \eprint{1410.3696}.

\bibitem[{\citenamefont{{Ackermann} et~al.}(2016)\citenamefont{{Ackermann},
  {Ajello}, {Albert}, {Atwood}, {Baldini}, {Ballet}, {Barbiellini}, {Bastieri},
  {Bechtol}, {Bellazzini} et~al.}}]{fermi_igrb_blazars}
\bibinfo{author}{\bibfnamefont{M.}~\bibnamefont{{Ackermann}}},
  \bibinfo{author}{\bibfnamefont{M.}~\bibnamefont{{Ajello}}},
  \bibinfo{author}{\bibfnamefont{A.}~\bibnamefont{{Albert}}},
  \bibinfo{author}{\bibfnamefont{W.~B.} \bibnamefont{{Atwood}}},
  \bibinfo{author}{\bibfnamefont{L.}~\bibnamefont{{Baldini}}},
  \bibinfo{author}{\bibfnamefont{J.}~\bibnamefont{{Ballet}}},
  \bibinfo{author}{\bibfnamefont{G.}~\bibnamefont{{Barbiellini}}},
  \bibinfo{author}{\bibfnamefont{D.}~\bibnamefont{{Bastieri}}},
  \bibinfo{author}{\bibfnamefont{K.}~\bibnamefont{{Bechtol}}},
  \bibinfo{author}{\bibfnamefont{R.}~\bibnamefont{{Bellazzini}}},
  \bibnamefont{et~al.}, \bibinfo{journal}{Physical Review Letters}
  \textbf{\bibinfo{volume}{116}}, \bibinfo{eid}{151105} (\bibinfo{year}{2016}),
  \eprint{1511.00693}.

\bibitem[{\citenamefont{{Neronov} et~al.}(2017)\citenamefont{{Neronov},
  {Semikoz}, and {Ptitsyna}}}]{blazars}
\bibinfo{author}{\bibfnamefont{A.}~\bibnamefont{{Neronov}}},
  \bibinfo{author}{\bibfnamefont{D.~V.} \bibnamefont{{Semikoz}}},
  \bibnamefont{and}
  \bibinfo{author}{\bibfnamefont{K.}~\bibnamefont{{Ptitsyna}}},
  \bibinfo{journal}{A\&A} \textbf{\bibinfo{volume}{603}}, \bibinfo{eid}{A135}
  (\bibinfo{year}{2017}), \eprint{1611.06338}.

\bibitem[{\citenamefont{{Aartsen}
  et~al.}(2017{\natexlab{b}})\citenamefont{{Aartsen}, {Abraham}, {Ackermann},
  {Adams}, {Aguilar}, {Ahlers}, {Ahrens}, {Altmann}, {Andeen}, {Anderson}
  et~al.}}]{blazars1}
\bibinfo{author}{\bibfnamefont{M.~G.} \bibnamefont{{Aartsen}}},
  \bibinfo{author}{\bibfnamefont{K.}~\bibnamefont{{Abraham}}},
  \bibinfo{author}{\bibfnamefont{M.}~\bibnamefont{{Ackermann}}},
  \bibinfo{author}{\bibfnamefont{J.}~\bibnamefont{{Adams}}},
  \bibinfo{author}{\bibfnamefont{J.~A.} \bibnamefont{{Aguilar}}},
  \bibinfo{author}{\bibfnamefont{M.}~\bibnamefont{{Ahlers}}},
  \bibinfo{author}{\bibfnamefont{M.}~\bibnamefont{{Ahrens}}},
  \bibinfo{author}{\bibfnamefont{D.}~\bibnamefont{{Altmann}}},
  \bibinfo{author}{\bibfnamefont{K.}~\bibnamefont{{Andeen}}},
  \bibinfo{author}{\bibfnamefont{T.}~\bibnamefont{{Anderson}}},
  \bibnamefont{et~al.}, \bibinfo{journal}{Ap.J.}
  \textbf{\bibinfo{volume}{835}}, \bibinfo{eid}{45}
  (\bibinfo{year}{2017}{\natexlab{b}}), \eprint{1611.03874}.

\bibitem[{\citenamefont{{Apel} et~al.}(2017)\citenamefont{{Apel},
  {Arteaga-Vel{\'a}zquez}, {Bekk}, {Bertaina}, {Bl{\"u}mer}, {Bozdog},
  {Brancus}, {Cantoni}, {Chiavassa}, {Cossavella} et~al.}}]{kascade}
\bibinfo{author}{\bibfnamefont{W.~D.} \bibnamefont{{Apel}}},
  \bibinfo{author}{\bibfnamefont{J.~C.} \bibnamefont{{Arteaga-Vel{\'a}zquez}}},
  \bibinfo{author}{\bibfnamefont{K.}~\bibnamefont{{Bekk}}},
  \bibinfo{author}{\bibfnamefont{M.}~\bibnamefont{{Bertaina}}},
  \bibinfo{author}{\bibfnamefont{J.}~\bibnamefont{{Bl{\"u}mer}}},
  \bibinfo{author}{\bibfnamefont{H.}~\bibnamefont{{Bozdog}}},
  \bibinfo{author}{\bibfnamefont{I.~M.} \bibnamefont{{Brancus}}},
  \bibinfo{author}{\bibfnamefont{E.}~\bibnamefont{{Cantoni}}},
  \bibinfo{author}{\bibfnamefont{A.}~\bibnamefont{{Chiavassa}}},
  \bibinfo{author}{\bibfnamefont{F.}~\bibnamefont{{Cossavella}}},
  \bibnamefont{et~al.}, \bibinfo{journal}{Ap.J.}
  \textbf{\bibinfo{volume}{848}}, \bibinfo{eid}{1} (\bibinfo{year}{2017}),
  \eprint{1710.02889}.

\bibitem[{\citenamefont{{Cha} et~al.}(1999)\citenamefont{{Cha}, {Sembach}, and
  {Danks}}}]{vela}
\bibinfo{author}{\bibfnamefont{A.~N.} \bibnamefont{{Cha}}},
  \bibinfo{author}{\bibfnamefont{K.~R.} \bibnamefont{{Sembach}}},
  \bibnamefont{and} \bibinfo{author}{\bibfnamefont{A.~C.}
  \bibnamefont{{Danks}}}, \bibinfo{journal}{Ap.J.Lett}
  \textbf{\bibinfo{volume}{515}}, \bibinfo{pages}{L25} (\bibinfo{year}{1999}),
  \eprint{astro-ph/9902230}.

\bibitem[{\citenamefont{Andersen et~al.}(2017)\citenamefont{Andersen,
  Kachelrie\ss, and Semikoz}}]{Andersen:2017yyg}
\bibinfo{author}{\bibfnamefont{K.~J.} \bibnamefont{Andersen}},
  \bibinfo{author}{\bibfnamefont{M.}~\bibnamefont{Kachelrie\ss}},
  \bibnamefont{and} \bibinfo{author}{\bibfnamefont{D.~V.}
  \bibnamefont{Semikoz}}, \bibinfo{journal}{arXiv:}
  \textbf{\bibinfo{volume}{1712.03153}} (\bibinfo{year}{2017}),
  \eprint{1712.03153}.

\bibitem[{\citenamefont{{Taylor} et~al.}(2014)\citenamefont{{Taylor}, {Gabici},
  and {Aharonian}}}]{halo_aharonian}
\bibinfo{author}{\bibfnamefont{A.~M.} \bibnamefont{{Taylor}}},
  \bibinfo{author}{\bibfnamefont{S.}~\bibnamefont{{Gabici}}}, \bibnamefont{and}
  \bibinfo{author}{\bibfnamefont{F.}~\bibnamefont{{Aharonian}}},
  \bibinfo{journal}{Phys. Rev. D} \textbf{\bibinfo{volume}{89}},
  \bibinfo{eid}{103003} (\bibinfo{year}{2014}), \eprint{1403.3206}.

\bibitem[{\citenamefont{Berezinsky et~al.}(1997)\citenamefont{Berezinsky,
  Kachelrie\ss, and Vilenkin}}]{Berezinsky:1997hy}
\bibinfo{author}{\bibfnamefont{V.}~\bibnamefont{Berezinsky}},
  \bibinfo{author}{\bibfnamefont{M.}~\bibnamefont{Kachelrie\ss}},
  \bibnamefont{and} \bibinfo{author}{\bibfnamefont{A.}~\bibnamefont{Vilenkin}},
  \bibinfo{journal}{Phys. Rev. Lett.} \textbf{\bibinfo{volume}{79}},
  \bibinfo{pages}{4302} (\bibinfo{year}{1997}), \eprint{astro-ph/9708217}.

\bibitem[{\citenamefont{Feldstein et~al.}(2013)\citenamefont{Feldstein,
  Kusenko, Matsumoto, and Yanagida}}]{Feldstein:2013kka}
\bibinfo{author}{\bibfnamefont{B.}~\bibnamefont{Feldstein}},
  \bibinfo{author}{\bibfnamefont{A.}~\bibnamefont{Kusenko}},
  \bibinfo{author}{\bibfnamefont{S.}~\bibnamefont{Matsumoto}},
  \bibnamefont{and} \bibinfo{author}{\bibfnamefont{T.~T.}
  \bibnamefont{Yanagida}}, \bibinfo{journal}{Phys. Rev.}
  \textbf{\bibinfo{volume}{D88}}, \bibinfo{pages}{015004}
  (\bibinfo{year}{2013}), \eprint{1303.7320}.

\bibitem[{\citenamefont{Esmaili and Serpico}(2013)}]{Esmaili:2013gha}
\bibinfo{author}{\bibfnamefont{A.}~\bibnamefont{Esmaili}} \bibnamefont{and}
  \bibinfo{author}{\bibfnamefont{P.~D.} \bibnamefont{Serpico}},
  \bibinfo{journal}{JCAP} \textbf{\bibinfo{volume}{1311}}, \bibinfo{pages}{054}
  (\bibinfo{year}{2013}), \eprint{1308.1105}.

\bibitem[{\citenamefont{{Murase} et~al.}(2015)\citenamefont{{Murase}, {Laha},
  {Ando}, and {Ahlers}}}]{murase}
\bibinfo{author}{\bibfnamefont{K.}~\bibnamefont{{Murase}}},
  \bibinfo{author}{\bibfnamefont{R.}~\bibnamefont{{Laha}}},
  \bibinfo{author}{\bibfnamefont{S.}~\bibnamefont{{Ando}}}, \bibnamefont{and}
  \bibinfo{author}{\bibfnamefont{M.}~\bibnamefont{{Ahlers}}},
  \bibinfo{journal}{Phys. Rev. Lett.} \textbf{\bibinfo{volume}{115}},
  \bibinfo{eid}{071301} (\bibinfo{year}{2015}), \eprint{1503.04663}.

\bibitem[{\citenamefont{Griest and Kamionkowski}(1990)}]{Griest:1989wd}
\bibinfo{author}{\bibfnamefont{K.}~\bibnamefont{Griest}} \bibnamefont{and}
  \bibinfo{author}{\bibfnamefont{M.}~\bibnamefont{Kamionkowski}},
  \bibinfo{journal}{Phys. Rev. Lett.} \textbf{\bibinfo{volume}{64}},
  \bibinfo{pages}{615} (\bibinfo{year}{1990}).

\bibitem[{\citenamefont{{IceCube-Gen2 Collaboration}
  et~al.}(2014)\citenamefont{{IceCube-Gen2 Collaboration}, {:}, {Aartsen},
  {Ackermann}, {Adams}, {Aguilar}, {Ahlers}, {Ahrens}, {Altmann}, {Anderson}
  et~al.}}]{gen2}
\bibinfo{author}{\bibnamefont{{IceCube-Gen2 Collaboration}}},
  \bibinfo{author}{\bibnamefont{{:}}}, \bibinfo{author}{\bibfnamefont{M.~G.}
  \bibnamefont{{Aartsen}}},
  \bibinfo{author}{\bibfnamefont{M.}~\bibnamefont{{Ackermann}}},
  \bibinfo{author}{\bibfnamefont{J.}~\bibnamefont{{Adams}}},
  \bibinfo{author}{\bibfnamefont{J.~A.} \bibnamefont{{Aguilar}}},
  \bibinfo{author}{\bibfnamefont{M.}~\bibnamefont{{Ahlers}}},
  \bibinfo{author}{\bibfnamefont{M.}~\bibnamefont{{Ahrens}}},
  \bibinfo{author}{\bibfnamefont{D.}~\bibnamefont{{Altmann}}},
  \bibinfo{author}{\bibfnamefont{T.}~\bibnamefont{{Anderson}}},
  \bibnamefont{et~al.}, \bibinfo{journal}{arXiv:}
  \textbf{\bibinfo{volume}{1412.5106}} (\bibinfo{year}{2014}),
  \eprint{1412.5106}.

\bibitem[{\citenamefont{{Adri{\'a}n-Mart{\'{\i}}nez}
  et~al.}(2016)\citenamefont{{Adri{\'a}n-Mart{\'{\i}}nez}, {Ageron},
  {Aharonian}, {Aiello}, {Albert}, {Ameli}, {Anassontzis}, {Andre},
  {Androulakis}, {Anghinolfi} et~al.}}]{km3net}
\bibinfo{author}{\bibfnamefont{S.}~\bibnamefont{{Adri{\'a}n-Mart{\'{\i}}nez}}},
  \bibinfo{author}{\bibfnamefont{M.}~\bibnamefont{{Ageron}}},
  \bibinfo{author}{\bibfnamefont{F.}~\bibnamefont{{Aharonian}}},
  \bibinfo{author}{\bibfnamefont{S.}~\bibnamefont{{Aiello}}},
  \bibinfo{author}{\bibfnamefont{A.}~\bibnamefont{{Albert}}},
  \bibinfo{author}{\bibfnamefont{F.}~\bibnamefont{{Ameli}}},
  \bibinfo{author}{\bibfnamefont{E.}~\bibnamefont{{Anassontzis}}},
  \bibinfo{author}{\bibfnamefont{M.}~\bibnamefont{{Andre}}},
  \bibinfo{author}{\bibfnamefont{G.}~\bibnamefont{{Androulakis}}},
  \bibinfo{author}{\bibfnamefont{M.}~\bibnamefont{{Anghinolfi}}},
  \bibnamefont{et~al.}, \bibinfo{journal}{Journal of Physics G Nuclear Physics}
  \textbf{\bibinfo{volume}{43}}, \bibinfo{eid}{084001} (\bibinfo{year}{2016}),
  \eprint{1601.07459}.

\bibitem[{\citenamefont{{Cui} et~al.}(2014)\citenamefont{{Cui}, {Liu}, {Liu},
  and {Ma}}}]{lhaaso}
\bibinfo{author}{\bibfnamefont{S.}~\bibnamefont{{Cui}}},
  \bibinfo{author}{\bibfnamefont{Y.}~\bibnamefont{{Liu}}},
  \bibinfo{author}{\bibfnamefont{Y.}~\bibnamefont{{Liu}}}, \bibnamefont{and}
  \bibinfo{author}{\bibfnamefont{X.}~\bibnamefont{{Ma}}},
  \bibinfo{journal}{Astroparticle Physics} \textbf{\bibinfo{volume}{54}},
  \bibinfo{pages}{86} (\bibinfo{year}{2014}).

\bibitem[{\citenamefont{Dzhappuev et~al.}(2017)}]{carpet}
\bibinfo{author}{\bibfnamefont{D.~D.} \bibnamefont{Dzhappuev}}
  \bibnamefont{et~al.}, \bibinfo{journal}{J. Phys. Conf. Ser.}
  \textbf{\bibinfo{volume}{934}}, \bibinfo{pages}{012022}
  (\bibinfo{year}{2017}).

\bibitem[{\citenamefont{{Aleksi{\'c}} et~al.}(2015)\citenamefont{{Aleksi{\'c}},
  {Ansoldi}, {Antonelli}, {Antoranz}, {Babic}, {Bangale}, {Barrio}, {Becerra
  Gonz{\'a}lez}, {Bednarek}, {Bernardini} et~al.}}]{crab}
\bibinfo{author}{\bibfnamefont{J.}~\bibnamefont{{Aleksi{\'c}}}},
  \bibinfo{author}{\bibfnamefont{S.}~\bibnamefont{{Ansoldi}}},
  \bibinfo{author}{\bibfnamefont{L.~A.} \bibnamefont{{Antonelli}}},
  \bibinfo{author}{\bibfnamefont{P.}~\bibnamefont{{Antoranz}}},
  \bibinfo{author}{\bibfnamefont{A.}~\bibnamefont{{Babic}}},
  \bibinfo{author}{\bibfnamefont{P.}~\bibnamefont{{Bangale}}},
  \bibinfo{author}{\bibfnamefont{J.~A.} \bibnamefont{{Barrio}}},
  \bibinfo{author}{\bibfnamefont{J.}~\bibnamefont{{Becerra Gonz{\'a}lez}}},
  \bibinfo{author}{\bibfnamefont{W.}~\bibnamefont{{Bednarek}}},
  \bibinfo{author}{\bibfnamefont{E.}~\bibnamefont{{Bernardini}}},
  \bibnamefont{et~al.}, \bibinfo{journal}{Journal of High Energy Astrophysics}
  \textbf{\bibinfo{volume}{5}}, \bibinfo{pages}{30} (\bibinfo{year}{2015}),
  \eprint{1406.6892}.

\bibitem[{\citenamefont{{Ackermann} et~al.}(2017)\citenamefont{{Ackermann},
  {Ajello}, {Baldini}, {Ballet}, {Barbiellini}, {Bastieri}, {Bellazzini},
  {Bissaldi}, {Bloom}, {Bonino} et~al.}}]{extended_galactic}
\bibinfo{author}{\bibfnamefont{M.}~\bibnamefont{{Ackermann}}},
  \bibinfo{author}{\bibfnamefont{M.}~\bibnamefont{{Ajello}}},
  \bibinfo{author}{\bibfnamefont{L.}~\bibnamefont{{Baldini}}},
  \bibinfo{author}{\bibfnamefont{J.}~\bibnamefont{{Ballet}}},
  \bibinfo{author}{\bibfnamefont{G.}~\bibnamefont{{Barbiellini}}},
  \bibinfo{author}{\bibfnamefont{D.}~\bibnamefont{{Bastieri}}},
  \bibinfo{author}{\bibfnamefont{R.}~\bibnamefont{{Bellazzini}}},
  \bibinfo{author}{\bibfnamefont{E.}~\bibnamefont{{Bissaldi}}},
  \bibinfo{author}{\bibfnamefont{E.~D.} \bibnamefont{{Bloom}}},
  \bibinfo{author}{\bibfnamefont{R.}~\bibnamefont{{Bonino}}},
  \bibnamefont{et~al.}, \bibinfo{journal}{Ap.J.}
  \textbf{\bibinfo{volume}{843}}, \bibinfo{eid}{139} (\bibinfo{year}{2017}).

\bibitem[{\citenamefont{{H.~E.~S.~S.~Collaboration}
  et~al.}(2016)\citenamefont{{H.~E.~S.~S.~Collaboration}, {Abdalla},
  {Abramowski}, {Aharonian}, {Ait Benkhali}, {Akhperjanian}, {Andersson},
  {Ang{\"u}ner}, {Arakawa}, {Arrieta} et~al.}}]{velajr}
\bibinfo{author}{\bibnamefont{{H.~E.~S.~S.~Collaboration}}},
  \bibinfo{author}{\bibfnamefont{H.}~\bibnamefont{{Abdalla}}},
  \bibinfo{author}{\bibfnamefont{A.}~\bibnamefont{{Abramowski}}},
  \bibinfo{author}{\bibfnamefont{F.}~\bibnamefont{{Aharonian}}},
  \bibinfo{author}{\bibfnamefont{F.}~\bibnamefont{{Ait Benkhali}}},
  \bibinfo{author}{\bibfnamefont{A.~G.} \bibnamefont{{Akhperjanian}}},
  \bibinfo{author}{\bibfnamefont{T.}~\bibnamefont{{Andersson}}},
  \bibinfo{author}{\bibfnamefont{E.~O.} \bibnamefont{{Ang{\"u}ner}}},
  \bibinfo{author}{\bibfnamefont{M.}~\bibnamefont{{Arakawa}}},
  \bibinfo{author}{\bibfnamefont{M.}~\bibnamefont{{Arrieta}}},
  \bibnamefont{et~al.}, \bibinfo{journal}{arXiv:}
  \textbf{\bibinfo{volume}{1611.01863}} (\bibinfo{year}{2016}),
  \eprint{1611.01863}.

\bibitem[{\citenamefont{{Abramowski} et~al.}(2012)\citenamefont{{Abramowski},
  {Acero}, {Aharonian}, {Akhperjanian}, {Anton}, {Balenderan}, {Balzer},
  {Barnacka}, {Becherini}, {Becker Tjus} et~al.}}]{velax}
\bibinfo{author}{\bibfnamefont{A.}~\bibnamefont{{Abramowski}}},
  \bibinfo{author}{\bibfnamefont{F.}~\bibnamefont{{Acero}}},
  \bibinfo{author}{\bibfnamefont{F.}~\bibnamefont{{Aharonian}}},
  \bibinfo{author}{\bibfnamefont{A.~G.} \bibnamefont{{Akhperjanian}}},
  \bibinfo{author}{\bibfnamefont{G.}~\bibnamefont{{Anton}}},
  \bibinfo{author}{\bibfnamefont{S.}~\bibnamefont{{Balenderan}}},
  \bibinfo{author}{\bibfnamefont{A.}~\bibnamefont{{Balzer}}},
  \bibinfo{author}{\bibfnamefont{A.}~\bibnamefont{{Barnacka}}},
  \bibinfo{author}{\bibfnamefont{Y.}~\bibnamefont{{Becherini}}},
  \bibinfo{author}{\bibfnamefont{J.}~\bibnamefont{{Becker Tjus}}},
  \bibnamefont{et~al.}, \bibinfo{journal}{A\&A} \textbf{\bibinfo{volume}{548}},
  \bibinfo{eid}{A38} (\bibinfo{year}{2012}), \eprint{1210.1359}.

\bibitem[{\citenamefont{{Aharonian}
  et~al.}(2006{\natexlab{a}})\citenamefont{{Aharonian}, {Akhperjanian},
  {Bazer-Bachi}, {Beilicke}, {Benbow}, {Berge}, {Bernl{\"o}hr}, {Boisson},
  {Bolz}, {Borrel} et~al.}}]{rxj1713}
\bibinfo{author}{\bibfnamefont{F.}~\bibnamefont{{Aharonian}}},
  \bibinfo{author}{\bibfnamefont{A.~G.} \bibnamefont{{Akhperjanian}}},
  \bibinfo{author}{\bibfnamefont{A.~R.} \bibnamefont{{Bazer-Bachi}}},
  \bibinfo{author}{\bibfnamefont{M.}~\bibnamefont{{Beilicke}}},
  \bibinfo{author}{\bibfnamefont{W.}~\bibnamefont{{Benbow}}},
  \bibinfo{author}{\bibfnamefont{D.}~\bibnamefont{{Berge}}},
  \bibinfo{author}{\bibfnamefont{K.}~\bibnamefont{{Bernl{\"o}hr}}},
  \bibinfo{author}{\bibfnamefont{C.}~\bibnamefont{{Boisson}}},
  \bibinfo{author}{\bibfnamefont{O.}~\bibnamefont{{Bolz}}},
  \bibinfo{author}{\bibfnamefont{V.}~\bibnamefont{{Borrel}}},
  \bibnamefont{et~al.}, \bibinfo{journal}{A\&A} \textbf{\bibinfo{volume}{449}},
  \bibinfo{pages}{223} (\bibinfo{year}{2006}{\natexlab{a}}),
  \eprint{astro-ph/0511678}.

\bibitem[{\citenamefont{{Aharonian}
  et~al.}(2006{\natexlab{b}})\citenamefont{{Aharonian}, {Akhperjanian},
  {Bazer-Bachi}, {Beilicke}, {Benbow}, {Berge}, {Bernl{\"o}hr}, {Boisson},
  {Bolz}, {Borrel} et~al.}}]{hess1825}
\bibinfo{author}{\bibfnamefont{F.}~\bibnamefont{{Aharonian}}},
  \bibinfo{author}{\bibfnamefont{A.~G.} \bibnamefont{{Akhperjanian}}},
  \bibinfo{author}{\bibfnamefont{A.~R.} \bibnamefont{{Bazer-Bachi}}},
  \bibinfo{author}{\bibfnamefont{M.}~\bibnamefont{{Beilicke}}},
  \bibinfo{author}{\bibfnamefont{W.}~\bibnamefont{{Benbow}}},
  \bibinfo{author}{\bibfnamefont{D.}~\bibnamefont{{Berge}}},
  \bibinfo{author}{\bibfnamefont{K.}~\bibnamefont{{Bernl{\"o}hr}}},
  \bibinfo{author}{\bibfnamefont{C.}~\bibnamefont{{Boisson}}},
  \bibinfo{author}{\bibfnamefont{O.}~\bibnamefont{{Bolz}}},
  \bibinfo{author}{\bibfnamefont{V.}~\bibnamefont{{Borrel}}},
  \bibnamefont{et~al.}, \bibinfo{journal}{A\&A} \textbf{\bibinfo{volume}{460}},
  \bibinfo{pages}{365} (\bibinfo{year}{2006}{\natexlab{b}}),
  \eprint{astro-ph/0607548}.

\bibitem[{\citenamefont{{Acero} et~al.}(2015)\citenamefont{{Acero},
  {Ackermann}, {Ajello}, {Albert}, {Atwood}, {Axelsson}, {Baldini}, {Ballet},
  {Barbiellini}, {Bastieri} et~al.}}]{fermi_cat}
\bibinfo{author}{\bibfnamefont{F.}~\bibnamefont{{Acero}}},
  \bibinfo{author}{\bibfnamefont{M.}~\bibnamefont{{Ackermann}}},
  \bibinfo{author}{\bibfnamefont{M.}~\bibnamefont{{Ajello}}},
  \bibinfo{author}{\bibfnamefont{A.}~\bibnamefont{{Albert}}},
  \bibinfo{author}{\bibfnamefont{W.~B.} \bibnamefont{{Atwood}}},
  \bibinfo{author}{\bibfnamefont{M.}~\bibnamefont{{Axelsson}}},
  \bibinfo{author}{\bibfnamefont{L.}~\bibnamefont{{Baldini}}},
  \bibinfo{author}{\bibfnamefont{J.}~\bibnamefont{{Ballet}}},
  \bibinfo{author}{\bibfnamefont{G.}~\bibnamefont{{Barbiellini}}},
  \bibinfo{author}{\bibfnamefont{D.}~\bibnamefont{{Bastieri}}},
  \bibnamefont{et~al.}, \bibinfo{journal}{Ap.J.Supp.}
  \textbf{\bibinfo{volume}{218}}, \bibinfo{eid}{23} (\bibinfo{year}{2015}),
  \eprint{1501.02003}.

\bibitem[{\citenamefont{Neronov et~al.}(2011)\citenamefont{Neronov, Semikoz,
  and Vovk}}]{Neronov:2010vv}
\bibinfo{author}{\bibfnamefont{A.}~\bibnamefont{Neronov}},
  \bibinfo{author}{\bibfnamefont{D.~V.} \bibnamefont{Semikoz}},
  \bibnamefont{and} \bibinfo{author}{\bibfnamefont{I.}~\bibnamefont{Vovk}},
  \bibinfo{journal}{Astron. Astrophys.} \textbf{\bibinfo{volume}{529}},
  \bibinfo{pages}{A59} (\bibinfo{year}{2011}), \eprint{1004.3767}.

\end{thebibliography}

\end{document}